\documentclass{article}

\usepackage{tikz}
\usepackage{amsmath}
\usepackage{amsfonts}
\usepackage{amssymb}
\usepackage{bm}
\usepackage{authblk}
\usepackage{accents}
\usepackage{bbm}
\usepackage{bm}
\usepackage{color}
\usepackage{graphicx}
\usepackage{amsthm}
\usepackage{enumerate}
\usepackage{dsfont}
\usepackage{mathrsfs}
\newlength{\dhatheight}

\usepackage{hyperref}
\usepackage{calligra}
\usepackage{relsize} 
\usepackage[thinlines]{easytable}
\begin{document}

\title{Thermodynamic bounds and symmetries in first-passage problems of  fluctuating currents }
\author[1]{Adarsh Raghu}
\author[1]{Izaak Neri}
\affil[1]{King's College London, London, United Kingdom}


\maketitle

\begin{abstract}
We develop a method for deriving thermodynamic bounds for first-passage problems of currents with two boundaries in Markov chains.  Using this method,  we derive a thermodynamic bound on the rate of dissipation in terms of the splitting probability and  the first-passage time statistics of a fluctuating current, which is a refinement of a previously derived inequality.   We also show that the concept of effective affinity, originally developed for continuous-time Markov chains, naturally extends to discrete-time Markov chains.  Furthermore,  we analyse symmetries in first-passage problems of fluctuating currents with two boundaries.   We  show that optimal currents — those for which the effective affinity fully accounts for the dissipation — satisfy a symmetry property: the current's average speed  to reach the positive threshold equals the current's speed to reach the negative threshold.   The developed approach uses a coarse-graining procedure for the average entropy production at random times and uses martingale methods to perform time-reversal of first-passage quantities.
\end{abstract}

\section{Introduction}

\subsection{Setup}
Nonequilibrium systems  can sustain currents with nonzero average rates, such as particle, energy, or  positional currents.   A central objective of stochastic thermodynamics is to incorporate fluctuations in a theory of nonequilibrium thermodynamics, see, e.g., Chapter XII of Ref.\cite{lifshitz}, and Refs.~\cite{Seki, peliti2021stochastic,limmer2024statistical,seifert}.
Current fluctuations  are conventionally studied at fixed times, e.g., with fluctuation relations~\cite{harris2007fluctuation}, or thermodynamic uncertainty relations~\cite{barato2015thermodynamic,gingrich2016dissipation}.  However, due to  noise,  in mesoscopic systems events of interest often take place at random times.   For such processes,   it is  more natural to study  fluctuations at  random termination times.

\begin{figure}[h!]{
    \centering
    \includegraphics[width=0.88\linewidth]{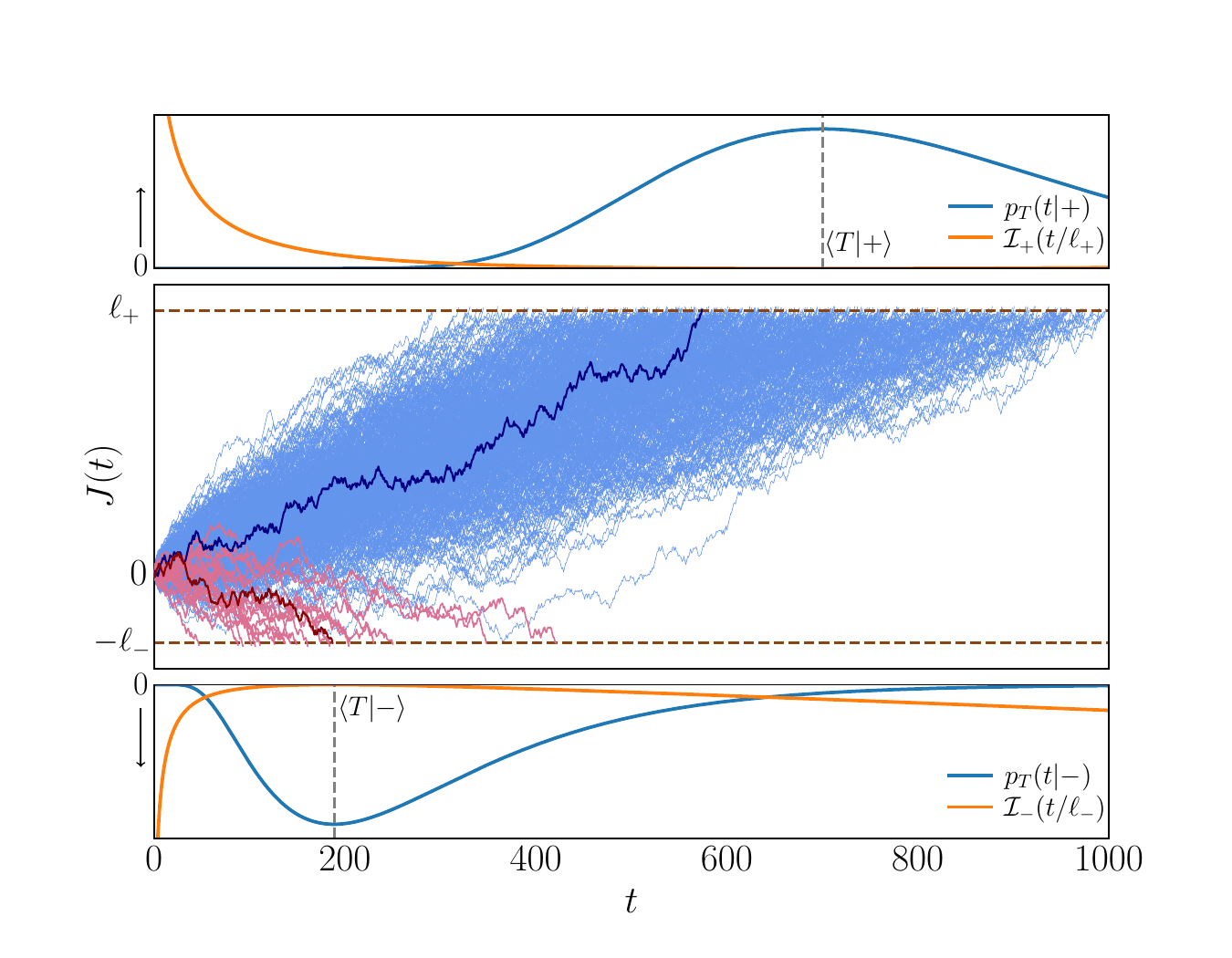}
    \caption{Graphical illustration of the first passage problem Eq.~\eqref{eq:T}. Multiple trajectories of a fluctuating  current $J(t)$  are plotted as a function of  time until the first time $T$ when $J$ exits the interval $(-\ell_-,\ell_+)$ (middle panel).  Trajectories   terminating at $\ell_+$ are coloured in blue and those terminating at $\ell_-$ are coloured in red.  One example trajectory for each case is highlighted in black.  The thresholds $\ell_\pm$ are marked as brown dashed lines.   The large deviation rate functions $\mathcal{I}_+(t/\ell_+)$ and  $\mathcal{I}_-(t/\ell_-)$ of the scaled first passage time conditioned on terminating at $\ell_+$ or $-\ell_-$, respectively, are plotted above and below the plot (yellow, solid lines); note that the y-axis is inverted in the bottom panel.  Furthermore, the corresponding conditioned first-passage time distributions $p_T(t|+)$ and $p_T(t|-)$ at both boundaries are shown (blue, solid lines).  Data is from the two-dimensional random walker model described in Sec.~\ref{sec:2DRandomd}, with parameters $\Delta=0.6$, $\rho=1$,  $\nu=\ln(4/3)$, and  thresholds  $\ell_+=100$ and $\ell_-=27$.}
    \label{fig:fig1}
}\end{figure}

In this paper we study current fluctuations through  first-passage problems of the form 
\begin{equation}
    T = {\rm inf}\left\{t \geq 0: J(t)\notin (-\ell_-,\ell_+)\right\},  \label{eq:T}
\end{equation} 
where $J(t)$ is a fluctuating current, 
where $T$ is the first time that the current exits a finite interval $(-\ell_-,\ell_+)$, and ``${\rm inf}$" is the infimum of a set; we use the convention   that $J(0)=0$.   The first-passage problems (\ref{eq:T}) generalise the gambler's ruin  problem to the  physically relevant case of   fluctuating currents.    Since currents are generally non-Markovian, analysing their first-passage statistics is more involved than in the gambler's ruin scenario. 
Key quantities  of interest  are  the distributions $p_T(t|-)$ and $p_T(t|+)$ of the first-passage time $T$ conditioned on $J(T)\geq \ell_+$ or   $J(T)\leq -\ell_-$, respectively, and   the splitting probability 
$p_-$, denoting the probability that $J(T)\leq -\ell_-$; see Fig.~\ref{fig:fig1} for a graphical illustration of this first-passage problem.    

\begin{figure}[h!]{
    \centering
\includegraphics[width=0.88\linewidth]{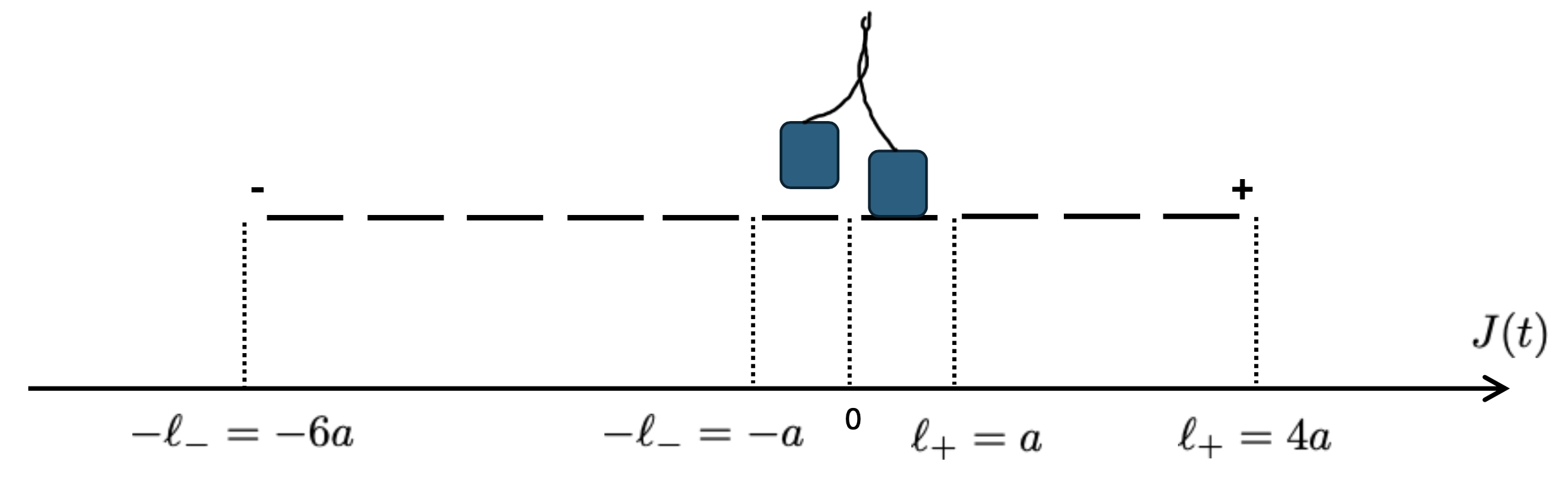}
       \caption{Illustrated application of the first-passage problem (\ref{eq:T}) for the case of a two-headed motor protein bounded to a biofilament.   The motion of the motor is biased towards the filament's plus end.  The fluctuating current $J(t)$ denotes the position of the center of mass of the motor along the filament.     In this example two natural first-passage problems $T$ are for $\ell_-=\ell_+=a$, in which case $T$ is the motor dwell time, and $-\ell_-$ and $\ell_+$ set equal to the end points of the filament, in which case  $T$    is the total duration of the motor's motion bound to the filament.  }\label{fig:motor}
}\end{figure}

First-passage problems of the form (\ref{eq:T}) model the timing of discrete events  in  nonequilibrium  mesoscopic systems.  A good example is the directed stepping of molecular motors along one-dimensional substrates, such as, motor proteins bound to biofilaments    (see Fig.~\ref{fig:motor} for an illustration). In this case,  $J$ is the positional current, denoting the position of the center of mass of the motor along the  biofilament.  The process $X$ represents  the internal coordinates of the motor, including the chemical states of the motor heads and the conformation of the motor~\cite{Liepelt, neri2022estimating, neri2023extreme}.  If we set  the thresholds $\ell_+=\ell_-=a$, with $a$ the length of a lattice site,  then  $J(T)>0$ and $J(T)<0$ correspond with the motor  stepping forwards or backwards, respectively.      Thus, in this case the  splitting probability $p_-$ is the probability for molecular motors to step backwards, and $T$ is the dwell time in between two steps~\cite{kolomeisky2005understanding} (see also \cite{fox2001rectified} for an alternative model).    Note that since the statistics of the  number of backward steps and the  dwell times  are experimentally measurable quantities of molecular motors ~\cite{nishiyama2002chemomechanical,asbury2003kinesin,rief2000myosin},  we can use them to   infer thermodynamic and kinetic properties  of molecular motors~\cite{seifert2019stochastic,neri2022estimating,nadal2025thermodynamic}, such as,  the efficiency of their chemomechanical coupling~\cite{kolomeisky2005understanding}.    Another natural choice for the thresholds $\ell_+$  and $\ell_-$ is to set them equal to the end points of the filament where the motor unbinds from the filament.  In this case $T$ denotes the total duration of the biased motion of the motor, and $p_-$ denotes a ``failure probability", the probability that the motor will detach from the filament without carrying its cargo to the correct end.   

 More complex processes, such as, cell-fate decisions~\cite{siggia2013decisions, desponds2020mechanism}, can also be modeled as first-passage problems of the type   (\ref{eq:T}).   Cell-fate decisions   involve the initiation of  gene transcription in response to extracellular ligand concentrations.  In this case,  the current  represents the accumulated evidence tracked by the system through a chemical reaction, and the sign of the current at the first-passage time corresponds with the decision taken by a system. 

In this paper, we address  in this paper two kind of problems related to first-passage times of the form (\ref{eq:T}).   The first concerns the relationship between the fluctuations of $T$ and the rate of dissipation $\dot{s}$ in the underlying process $X$. The second involves the symmetry properties of the statistics of $T$ at positive  versus negative thresholds. Below, we provide a brief overview of these two problems, followed by a discussion of the novel contributions presented in this paper.

\subsection{Trade-off relations between speed, accuracy, and dissipation}

A generic feature of first-passage problems is the trade-off between speed, quantified by the mean first-passage time $\langle T \rangle$, and accuracy, quantified by the splitting probability $p_-$.   For instance, increasing the thresholds $\ell_-$ and $\ell_+$, the  mean first-passage time $\langle T\rangle$  increases and the splitting probability $p_-$ decreases.    This is the speed-accuracy  trade-off that is observed in a large number of decision making systems, see Refs.~\cite{chittka2009speed,bogacz2010neural} for reviews. 

A natural question for nonequilibrium thermodynamics is how dissipation, quantified by the entropy production rate $\dot{s}$, influences the speed-accuracy trade-off in first-passage problems.   A couple of trade-off relations between speed, accuracy, and dissipation  have appeared before in the  literature, and we review them below.  These relations have  been derived  in the general setup of  fluctuating currents in continuous-time Markov chains and in the limit of large thresholds. We assume in what follows, without loss of generality, that  the fluctuating current $J(t)$ is increasing on average.  

References~\cite{decision1,neri2022universal,raghu2024effective} show that the following first-passage trade-off relation,
\begin{equation}
\dot{s} \geq \frac{\ell_+}{\ell_-} \frac{|\ln p_-|}{\langle T\rangle}(1+o_{\ell_{\rm  min}}(1)) ,\label{eq:unc}
\end{equation} 
holds for arbitrary currents, 
where the little-o notation $o_{\ell_{\rm  min}}(1)$ represents a function that converges to zero when $\ell_{\rm min} = {\rm min}\left\{\ell_+,\ell_-\right\}$ diverges.   The inequality (\ref{eq:unc}) firmly establishes the trade-off between speed, accuracy, and dissipation, as quantified by $\langle T \rangle$, $|\ln p_-|$, and $\dot{s}$, respectively.

Applied to molecular motors, the inequality (\ref{eq:unc})  relates  probability of backward steps ($p_-$)  and the mean dwell time ($\langle T\rangle$), to the rate of dissipation ($\dot{s}$).   In this context, the splitting probability   $p_-$ is a natural  measure of (in)accuracy of molecular motors, as  backward steps can be considered errors due to thermal fluctuations in which the motor moves in the opposite direction to its intended trajectory.   
The inequality~(\ref{eq:unc}) can be used to infer the efficiency of the chemomechanical coupling in molecular motors from the observation of the statistics of backward steps~\cite{kolomeisky2005understanding}.   For example,  using the statistics of Kinesin-1 backward steps,  Ref.~\cite{neri2022estimating} estimated that Kinesin-1 converts Gibbs free energy into mechanical work with an efficiency that is smaller than $0.6$.

An equivalent form of the inequality (\ref{eq:unc}) is 
\begin{equation}
\dot{s} \geq \overline{j}a^\ast,\label{eq:sjas} 
\end{equation} 
where 

\begin{equation}
\overline{j} = \lim_{t\to\infty} \frac{J(t)}{t} 
\end{equation}
is the average rate of increase of the current $J(t)$ and
\begin{equation} 
\lim_{\ell_-\rightarrow \infty}\frac{|\ln p_-|}{\ell_-}  = a^\ast  \label{eq:thermoSplit}
\end{equation}
is the exponential decay constant of the splitting probability~\cite{raghu2024effective}; (\ref{eq:unc}) and (\ref{eq:sjas}) are equivalent in the sense that the right-hand sides of the two equalities are identical, which can be seen from Eq.\eqref{eq:thermoSplit} and the fact that $ \lim_{\ell_{\rm min} \to \infty} \frac{\ell_+}{\langle T\rangle}=\overline{j}$.   
Interestingly, in systems with coupled currents, the decay constant $a^\ast$ exhibits properties similar to those of chemical affinities in uncoupled systems~\cite{raghu2024effective}, and therefore has been coined the  {\it effective affinity}, see  Refs.~\cite{polletini1,polettini2019effective, neri2023extreme, raghu2024effective}.   Notably the effective affinity  determines the direction of the current ($\overline{j}a^\ast\geq 0$); captures a portion of the total  amount of dissipation $\dot{s}$ [see Eq.~(\ref{eq:sjas})]; and it determines the  statistics of the fluctuations of the currents against its average flow [see Eq.~(\ref{eq:thermoSplit})].    For systems with uncoupled currents $a^\ast$ equals the chemical affinity~\cite{de1936thermodynamic,de2013non}, and for currents that are proportional to the fluctuating entropy production, or more broadly belong to the same cycle equivalence class as the fluctuating entropy production~\cite{raghu2024effective}, $a^\ast = \dot{s}/\overline{j}$, so that the equality in  (\ref{eq:sjas}) is attained.

An alternative trade-off relation  between speed, accuracy, and dissipation is the thermodynamic uncertainty relation for first-passage times times~\cite{gingrich2017fundamenta},  given by 
\begin{equation}
\dot{s}\geq \frac{2}{\langle T\rangle_+} \frac{\langle T\rangle^2_+}{\langle T^2\rangle_+-\langle T\rangle^2_+}(1+o_{\ell_+}(1)), \label{eq:TUR}
\end{equation} 
where here and throughout the paper, we use the notation $\langle\cdot\rangle_+ = \langle\cdot\vert J(T)\geq \ell_+\rangle$ and $\langle\cdot\rangle_- = \langle\cdot\vert J(T)\leq -\ell_-\rangle$ to indicate averages conditioned on the first passage problem \eqref{eq:T} terminating at the positive and negative threshold, respectively.   The main distinction between the inequalities (\ref{eq:unc}) and (\ref{eq:TUR}) is how accuracy in $J$ is quantified, with the uncertainy relation using the   inverse Fano Factor $\langle T\rangle^2_+/(\langle T^2\rangle_+-\langle T\rangle^2_+)$ and the first-passage trade-off relation using  the splitting probability $p_-$.

In the linear response regime, where the fluctuations are Gaussian, the first-passage trade-off relation (\ref{eq:unc}) and the thermodynamic uncertainty relation (\ref{eq:TUR}) are equivalent, as the right-hand sides of the inequalities (\ref{eq:unc}) and (\ref{eq:TUR}) are identical in this limit.   Instead, far from thermal equilibrium, where fluctuations are not gaussian, these two inequalities can  differ significantly.  Examples show that the  first-passage trade-off captures a larger portion of dissipation far from thermal equilibrium than the thermodynamic uncertainty relation, as the right-hand side of (\ref{eq:unc}) scales proportionally with $\dot{s}$, while the right-hand side (\ref{eq:TUR})  does not~\cite{neri2022estimating}.  This can be understood from the fact that the splitting probability $p_-$ quantifies atypical fluctuations in the process $J$, while the Fano factor  quantifies  typical fluctuations of $J$.

Another inequality    that is discussed in the literature is  the time-dissipation uncertainty relation~\cite{falasco2020dissipation,yan2022experimental}
\begin{equation}
\dot{s} \geq \frac{1}{\langle T\rangle_+} (1+o_{\ell_+}(1)).   \label{eq:TimeDiss}
\end{equation}
However, in the present context of first-passage times of time-additive observables, Eq.~(\ref{eq:TimeDiss})  is  vacuous.  Indeed, comparing (\ref{eq:TimeDiss})  with the inequalities (\ref{eq:unc}) and (\ref{eq:TUR}), one notices that the right-hand side of (\ref{eq:TimeDiss}) misses an infinitely large prefactor that quantifies the uncertainty in the process [this is $\langle T\rangle^2_+/(\langle T^2\rangle_+-\langle T\rangle^2_+)$ in the case of  (\ref{eq:TUR}) and $|\ln p_-|$ in the case of (\ref{eq:unc})].   This prefactor is necessary. Indeed, as    $\langle T\rangle_+\sim \ell_+$,  in the limit of large $\ell_+$ the  time-dissipation uncertainty relation (\ref{eq:TimeDiss}) is equivalent with the second law of thermodynamics $\dot{s}\geq 0$.    Instead, the inequalities  (\ref{eq:unc}) and (\ref{eq:TUR}) have right-hand sides that are strictly positive in this  asymptotic limit, and  these inequalities are thus  not equivalent with the second law of thermodynamics.         

Thermodynamic bounds have also been derived for the distributions of first-passage times $T$ in the limit of large thresholds.    
In this limit, the first-passage time $T$ satisfies a large-deviation principle at both thresholds with speeds $\ell_+$ and $\ell_-$~\cite{gingrich2017fundamenta,neri2025martingale}, respectively.  The corresponding rate functions are defined by 
\begin{equation}
\mathcal{I}_+(\tau) =-\lim_{\ell_+\rightarrow \infty} \frac{\ln p_{T}(\ell_+\tau |+)}{\ell_+} \quad  {\rm and} \quad \mathcal{I}_-(\tau) =-\lim_{\ell_-\rightarrow \infty} \frac{\ln p_{T}(\ell_-\tau |-)}{\ell_-} ,  \label{eq:rate}
\end{equation}
where $\tau \geq 0$, and where $p_{T}(\cdot\vert +)$ or $p_{T}(\cdot\vert -)$ denote the distributions of the first-passage time $T$, conditioned on the current $J$ exiting the interval $(-\ell_-, \ell_+)$ for the first time through the thresholds $\ell_+$ or $-\ell_-$, respectively.  Note that the argument $\tau$ of the rate function is the time  rescaled by the speed of the large deviation principle. Note that the  large deviation principles at both thresholds hold independently of each other.
The scaled cumulant generating functions 
\begin{equation}
m_+(\mu) =\lim_{\ell_+\rightarrow \infty} \frac{1}{\ell_+}\langle e^{\mu T}\rangle_+  \quad {\rm and} \quad  m_-(\mu) =\lim_{\ell_-\rightarrow \infty} \frac{1}{\ell_-}\langle e^{\mu T}\rangle_-  \label{eq:mDef}
\end{equation}
are the Fenchel-Legendre transforms of the rate functions $\mathcal{I}_+$ and $\mathcal{I}_-$, respectively, viz.,
\begin{equation}
m_+(\mu) = {\rm sup}_{\tau\in \mathbb{R}^+}(\tau\mu -\mathcal{I}_+(\tau)) \quad {\rm and} \quad  m_-(\mu) ={\rm sup}_{\tau\in \mathbb{R}^+}(\tau\mu -\mathcal{I}_-(\tau)).  \label{eq:mlegendre}
\end{equation}
Using large deviation theory and martingale theory, it was shown that  scaled cumulant generating functions satisfy the inequalities~\cite{gingrich2017fundamenta}
\begin{equation}
m_+(\mu)  \geq  \frac{\dot{s}}{2\overline{j}} \left(1-\sqrt{1-4\mu/\dot{s}}\right) \label{eq:mPIneq}
\end{equation}
and \cite{neri2025martingale}
\begin{equation}
m_-(\mu)  \geq  \frac{\dot{s}}{2\overline{j}} \left(1-\sqrt{1-4\mu/\dot{s}}\right) + a^\ast - \frac{\dot{s}}{\overline{j}}. \label{eq:mMIneq}
\end{equation}
The first inequality yields the thermodynamic uncertainty relation    for first-passage times, Eq.~(\ref{eq:TUR}), 
while, excluding special cases,  a similar inequality does not apply at the negative threshold~\cite{neri2025martingale}.

Lastly, let us mention that other  trade-off relations  have been discussed in the literature~\cite{garrahan2017simple, pal2021thermodynamic, hiura2021kinetic, wampler2021skewness, liu2024semi, bakewell2024bounds, macieszczak2024ultimate}.   Some of these are on  first-passage problems in semi-Markov processes, which is relevant for problems in quantum mechanics,  or  on first-passage problems of time-additive observables that are not fluctuating currents.  Although these results are  interesting, they are not directly related to the problems discussed in this paper, and thus we will not discuss them further.
  
\subsection{First-passage time symmetries}
 We say that $T$ satisfies a first-passage symmetry when the  statistics of first-passage times at the negative threshold equal those at the positive threshold, i.e., 
\begin{equation}
\mathcal{I}_-(\tau)=\mathcal{I} _+(\tau)\label{eq:opt}
\end{equation}  
for all $\tau\geq 0$.

While the trade-off relations (\ref{eq:thermoSplit}) and (\ref{eq:TUR}) hold for arbitrary currents, the symmetry relation (\ref{eq:opt}) is more specific. Thus, by identifying the currents to which the symmetry (\ref{eq:opt}) applies, one can infer kinetic properties of mesoscopic systems upon observing this symmetry.

The first-passage symmetry (\ref{eq:opt}) was first observed in  Pascal's gambler’s ruin problem, see e.g.,~Ref.~\cite{samuels1975classical}. In physics, interest in this symmetry increased after it was empirically observed in the dwell times  between consecutive steps of kinesin motors~\cite{nishiyama2002chemomechanical}, an observation that is also interesting for describing time-reversibility in coarse-grained models of molecular motors~\cite{wang2007detailed}.
References~\cite{kolomeisky2005understanding, qian2006generalized, ge2008waiting,piephoff2025first} demonstrated that Eq.~(\ref{eq:opt}) applies to currents in unicyclic systems, while Refs.~\cite{neri2017statistics, krapivsky2018first} derived an equivalent symmetry for one-dimensional overdamped Langevin processes with constant drift.
Given the link between dissipation and time-reversibility, one might expect a relation between first-passage time symmetries and entropy production. This connection was confirmed in Refs.~\cite{saito2016waiting, decision1, neri2017statistics, roldan2022martingales}, which showed that the symmetry (\ref{eq:opt}) holds for currents proportional to the fluctuating entropy production. This result was further extended in Ref.~\cite{raghu2024effective} to include currents that are cycle-equivalent to  the fluctuating entropy production.    Two currents are cycle equivalent if they have the same total increment  along a set of fundamental cycles in the graph.   Therefore,  in the case of unicyclic systems, all currents are cycle equivalent up to a proportionality constant. 
Other symmetries related to Eq.~(\ref{eq:opt})  include symmetry relations for cycle formation times~\cite{jia2016cycle} and for first-passage times of winding numbers~\cite{bauer2014affinity}.

The symmetry relation~(\ref{eq:opt}) follows from the Gallavotti-Cohen-type symmetry~\cite{lebowitz1999gallavotti}, 
\begin{equation}
\mathcal{I}_J(j) = \mathcal{I}_J(-j)- a^\ast j, \label{eq:GCsymLDR}
\end{equation}
where $\mathcal{I}_J$ is the (large deviation) rate function of the current $J(t)$ in the limit of large $t$, and $a^\ast$ is the effective affinity.   Instead, for currents that do not satisfy the Gallavotti-Cohen-type fluctuation relation,  Eq.~(\ref{eq:opt}) does  in general not apply~\cite{neri2025martingale, sarmiento2025first}.

Currents that are cycle-equivalent with the stochastic entropy production are also optimal currents, in the sense that the equality in (\ref{eq:sjas}) is attained.   Therefore, such currents satisfy both (\ref{eq:sjas})  and the symmetry relation (\ref{eq:opt}).
This suggests that there exists a relationship between optimality and symmetry.     However, it should be noted that the symmetry relation (\ref{eq:opt}) is not a sufficient condition for current optimality, and it remains an open question whether  it is a necessary condition.  Accordingly, the precise relationship between symmetry and dissipation remains an open question, and this paper offers new insights into this issue. 

Since the symmetry relation (\ref{eq:opt}) does  in general not apply, one may wonder whether first-passage times of generic currents satisfy a generalised symmetry.
For edge currents, which are a specific class of currents that count the net number of transitions between two states in a Markov process, the following  generalised symmetry relation  
 \begin{equation}
\mathcal{I}_-(\tau) = \hat{\mathcal{I}}^\dagger_+(\tau) \label{eq:generalisedSymmetry2}
\end{equation}
applies~\cite{polletini1}, 
where $\hat{\mathcal{I}}^\dagger_+(\tau)$ is the rate function of $T$ at the positive threshold in a conjugate process.      The first-passage symmetry  (\ref{eq:generalisedSymmetry2}) follows from  the generalised fluctuation symmetry
\begin{equation}
\mathcal{I}_J(j) = \hat{\mathcal{I}}^\dagger_J(-j) - a^\ast j, \label{eq:GFT}
\end{equation}
which has also been derived in the specific case of edge currents~\cite{polletini1} .   In (\ref{eq:GFT}), 
   $\hat{\mathcal{I}}^\dagger_J$ is the rate function of $J(t)$ in a conjugate process defined by the effective affinity $a^\ast$.

   \subsection{Results of this Paper}
  In this Paper we expand on the results above. We develop a  method to derive the inequalities  (\ref{eq:unc})  and (\ref{eq:sjas})   that is different from derivations presented previously in the literature~\cite{neri2022universal,raghu2024effective}.  This new approach lends itself naturally to extensions, and we use this approach  to extend the inequalities  (\ref{eq:sjas}) and  (\ref{eq:unc}) in two  meaningful ways.  

First, we show that the inequalities   (\ref{eq:unc})  and (\ref{eq:sjas})    also apply  for Markov chains in discrete time;  this stands in contrast with the inequalities (\ref{eq:TUR})-(\ref{eq:mMIneq}) that do not extend to the discrete-time setups~\cite{proesmans2017discrete}.    Hence, the effective affinity $a^\ast$ is also meaningful as an affinity-like quantity of coupled currents in discrete-time Markov chains.  

Second, we derive the refined inequality 
  \begin{equation}
\dot{s} \geq   \frac{\ell_+}{\langle T\rangle} \left(\frac{|\ln p_- |}{\ell_-} + \mathcal{I}_-\left(1/\overline{j}\right)\right)(1+o_{\ell_{\rm  min}}(1)).   \label{eq:totalbound}
 \end{equation}   
The difference between the  inequalities  (\ref{eq:totalbound}) and (\ref{eq:unc})   is that the right-hand side of Eq.~(\ref{eq:totalbound}) also depends on the fluctuation properties of $T$ through the term  $\mathcal{I}_-(1/\overline{j})$.     Since the term $\mathcal{I}_-(1/\overline{j})>0$ the inequality (\ref{eq:totalbound}) implies the inequality (\ref{eq:unc}).

Just as the  first-passage inequality (\ref{eq:sjas}) can be expressed as the inequality (\ref{eq:unc}) that involves  the fluctuations of $J(t)$ through  $a^\ast$, we can express the inequality (\ref{eq:totalbound}) in terms of current fluctuations at fixed time.  Specifically,  the inequality (\ref{eq:totalbound})  is equivalent to 
\begin{equation}
    \dot{s} \geq  I_J(-\overline{j}) = \tilde{a} \overline{j} - \lambda_J(\tilde{a}),\label{eq:newboundIJform}
\end{equation}
where $\tilde{a}$ is the positive value of $a$ such that $\lambda'_J(\tilde{a}) = \overline{j}>0$, and with the prime indicating a derivative towards $a$.  If $\tilde{a}=a^\ast$, then  the right-hand side of (\ref{eq:newboundIJform}) equals the right-hand side of (\ref{eq:sjas}).

The inequality (\ref{eq:totalbound}) clarifies the  anticipated relation between symmetry and optimality (optimality in the sense of attaining the equality in (\ref{eq:sjas})).   From the bound   (\ref{eq:totalbound})  it follows that optimal currents satisfy the symmetry relation  
\begin{equation}
\lim_{\ell_-\rightarrow \infty}\frac{\langle T\rangle_-}{\ell_-} = \lim_{\ell_+\rightarrow \infty}\frac{\langle T\rangle_+}{\ell_+}, \label{eq:weaksymm}
\end{equation}
i.e., the speed of reaching the positive threshold equals the speed of reaching the negative threshold.  However, the converse is not true, i.e., the symmetry relation (\ref{eq:weaksymm}) does not imply optimality.  Note that  the   symmetry relation (\ref{eq:weaksymm}) is   equivalent to the condition $a^\ast = \tilde{a}$, and thus general        
currents do not satisfy the symmetry relation (\ref{eq:weaksymm}), see~Ref.~\cite{neri2025martingale}.

Furthermore,  we show that all fluctuating currents satisfy a generalised symmetry relation   of the form (\ref{eq:generalisedSymmetry2}) for an appropriately defined conjugate process.      The conjugate process  is  the time-reversal of the dual process identified in Ref.~\cite{neri2025martingale}, which is a  
a tilted Markov process~\cite{chetrite_nonequilibrium_2015}  with the tilting parameter equal to the  effective affinity $a^\ast$.

From a methodological point of view, we  extend techniques known at fixed time to random times.    Specifically, we develop a  method for coarse-graining the average entropy production at  first-passage times, $\langle S(T)\rangle$, which generalises existing methods for coarse-graining the average entropy production at fixed times, $\langle S(t)\rangle$ (see,  e.g., Refs.~\cite{coarseGraining,coarseGraining2, harunari_what_2022,partial,blom2024milestoning,ertel2024estimator,fritz2025entropy}).   A key insight is that $\langle S(T)\rangle$ can be expressed as a Kullback-Leibler divergence between two probability distributions.   Those probability distributions are defined on an unusual set consisting of all possible realisations of the process up to the first-passage time. By applying standard coarse-graining methods to this Kullback-Leibler divergence, we derive bounds on the rate of dissipation. However, doing so a complication arises, as the resultant  bounds involve probabilities of first-passage events in the time-reversed dynamics, which are generally of less practical interest. Fortunately, using martingale theory, we can relate the first-passage quantities in the time-reversed dynamics to those in the forward dynamics~\cite{raghu2024effective,neri2025martingale}.  

\subsection{Outline}
The paper is structured as follows: In Sec.~\ref{sec:setup} we introduce the system setup.   In Sec.~\ref{sec:firstbound} we derive the inequalities (\ref{eq:sjas}) and  (\ref{eq:unc})  with an approach that is different from derivations previously considered in the literature~\cite{neri2022universal,raghu2024effective}.  Furthermore, we show that these inequalities apply to both continuous-time and discrete-time Markov chains.   In Sec.~\ref{sec:PT}  we derive the bounds (\ref{eq:totalbound}) and  (\ref{eq:newboundIJform}).   In Sec.~\ref{sec:symmetry} we discuss the relation between current symmetry and optimality, and in Sec.~\ref{sec:FT} we derive the generalised fluctuation relation (\ref{eq:generalisedSymmetry2}) for  generic currents.  Section~\ref{sec:examples} illustrates the results on two simple examples of a Markov jump process.  We end the paper with a discussion in Sec.~\ref{sec:disc} and a few appendices with technical details. 
 
\section{System setup: fluctuating currents and  entropy production  in  Markov chains} \label{sec:setup}
We define the system setup that we use for studying  first-passage times of  currents.   This setup consists of fluctuating currents in stationary Markov chains.   Subsequently, we define entropy production in such Markov chains, relating the setup to  nonequilibrium thermodynamics. The results we derive in this paper are valid within   the setup described below, and no additional assumptions are required.

\subsection{Markov chains}
We consider a Markov process $X(t)\in \mathcal{X}$ with $\mathcal{X}$ a finite set and with $t\in \mathbb{I}$ a time index that can be continuous $\mathbb{I} = \mathbb{R}^+$ or discrete $\mathbb{I}= \mathbb{N}$. Without loss of generality we can set $\mathcal{X} = \left\{1,2,\ldots,|\mathcal{X}|\right\}$.

\subsubsection{Discrete time.}\label{sec:Discrete}
In discrete time, the  sample space $\Omega$   consists of all trajectories of the form 
\begin{equation}
x^\infty_0 = (x(0),x(1),x(2),\ldots),
\end{equation} 
where $x(t)\in \mathcal{X}$ for all $t\geq0$.   

The random variable $X(t)$ is a map from  $\Omega$ to $\mathcal{X}$ so that  
$X(t,x^\infty_0) = x(t)$  returns the value of the trajectory $x^\infty_0$ at the time $t$.  Let $f(X^t_{0})$ be a  function of the trajectory $X^t_0 = (X(0),X(1),\ldots,X(t))$.  The  average  value of $f$ is given by 
\begin{equation}
\langle f(X^t_{0}) \rangle = \sum_{x^t_0\in \mathcal{X}^{t+1}} \mathbb{P}\left(X^t_{0}=x^t_0\right) f(x^t_0), \label{eq:aveargef} 
\end{equation}
where $\mathbb{P}\left(X^t_{0}=x^t_0\right)$ is the probability to observe the  trajectory $x^t_0$.  

For a time-homogeneous Markov chain,  $\mathbb{P}$   is specified by the probability mass function of the  initial configuration
\begin{equation}
p_{\rm init} (x) = \mathbb{P}\left(X(0)=x\right),
\end{equation}
and the matrix of transition probabilities 
\begin{equation}
\mathbf{q}_{yx} = \mathbb{P}\left(X(t)=x|X(t-1)=y\right).
\end{equation}
It follows from the Markov property that 
\begin{equation}
\mathbb{P}(X^t_0 = x^t_0) = p_{\rm init} (x_0) \prod^{t}_{t'=1}\mathbf{q}_{x(t'-1)x(t')} = p(x^t_0), \label{eq:pPathDiscrete}
\end{equation}
 where we have introduced the simpler notation $p(x^t_0)$ for $\mathbb{P}(X^t_0 = x^t_0)$.      

The transition matrix $\mathbf{q}_{xy}$ satisfies $\sum_{y\in \mathcal{X}}\mathbf{q}_{xy} = 1$.   We make two additional assumptions on $\mathbf{q}$.    We assume that  the graph of admissible transitions is strongly connected, and we assume that $\mathbf{q}_{xy}>0$ implies that $\mathbf{q}_{yx}>0$, so  that $X$ is both ergodic and  time-reversible.   
Under these assumptions, there exists a unique probability mass function   $p_{\rm ss}(x)$ satisfying $\sum_{x\in \mathcal{X}}p_{\rm ss}(x)\mathbf{q}_{xy} = p_{\rm ss}(y)$. The  distribution $p_{\rm ss}(y)$  is  the stationary distribution, which specifies the probability of $X(t)=y$ in the limit of large $t$.  For the derivation of this paper's main results we work with stationary Markov chains for which  $p_{\rm init} = p_{\rm ss}$. 

\subsubsection{Continuous time.} In continuous time, the sample space $\Omega$ consists of all trajectories
\begin{equation}
x^{\infty}_0 = \left\{x(t):t\in\mathbb{R}^+\right\},
\end{equation}
where $x(t)$ is a right-continuous (i.e., $\lim_{\epsilon\rightarrow 0^+}x(t+\epsilon) = x(t)$) and piecewise constant function from $\mathbb{R}^+$ to $\mathcal{X}$.     As the function is piecewise constant and right-continuous, the  trajectories $x^{\infty}_0$ are specified by a discrete-time trajectory  $(y(0),y(1),y(2)\ldots)$, with $y(j)\in \mathcal{X}$,  and an increasing sequence of jump times $(t(0),t(1),t(2),\ldots)$ with $t(0)=0$, so that $x(t) = y({\rm max}\left\{t(j): j\in \mathbb{N} , t(j) \leq t\right\})$~(see Chapter 3 of \cite{norris1998markov}).   We write averages of functionals $f(X^t_0)$  as in Eq.~(\ref{eq:aveargef}), but now the sum has to be understood  over all possible right-continuous trajectories $x^t_0$; this involves summing over all possible number of jumps $n$, summing over all possible sequences of states $y^{n}_0$, and integrating over all possible jump times $t^{n}_0$ with $t(n)\leq t$.      

The probability measure $\mathbb{P}$ is in this case  specified by the probability mass function 
\begin{equation}
p_{\rm init} (x) = \mathbb{P}\left(X(0)=x\right),
\end{equation}
and the transition rate matrix 
\begin{equation}
\mathbf{q}_{yx} = \lim_{{\rm d}t\rightarrow 0} \frac{\mathbb{P}\left(X(t+{\rm d}t)=x|X(t)=y\right)}{{\rm d}t}
\end{equation}
for all $x\neq y$; we set the diagonal elements of  $\mathbf{q}$  equal to $\mathbf{q}_{xx} = -\sum_{y\in \mathcal{X}}\mathbf{q}_{xy}$.   
Using the Markov property we can assign the following probabilistic weight to a trajectory $x^t_0$,  
\begin{eqnarray}
 \mathbb{P}\left(X^t_0=x^t_0\right) &=& p_{\rm init}(y(0))e^{\mathbf{q}_{y(n)y(n)} (t-t(n))}
\nonumber\\
&\times& \prod^{n}_{j=1} \left\{{\rm d}[t(j)-t(j-1)] \:  \, e^{\mathbf{q}_{y(j)y(j)} [t(j)-t(j-1)]} \mathbf{q}_{y(j-1)y(j)}\right\}=:p(x^t_0),\label{eq:PathContinuous}
\end{eqnarray}
where $n = n(x^t_0)$ is the number of jumps in the trajectory $x^t_0$.    

As in discrete time, we assume that the graph of admissible transitions is strongly connected and we assume that  $\mathbf{q}_{xy}>0$ if and only if  $\mathbf{q}_{yx}>0$, so that $X$ is ergodic and time-reversible.  We work in the stationary regime for which $p_{\rm init} = p_{\rm ss}$.

\subsection{Fluctuating currents}
A fluctuating current $J(t)$ is  a time-additive observable of the form 
\begin{equation}
J(t) = \sum_{x\in \mathcal{X}}\sum_{y\in \mathcal{X}\setminus \left\{x\right\}}c_{xy}N^{xy}(t), \label{eq:J}
\end{equation}
with $N^{xy}(t)$ the number of times that the process $X$ has jumped from $x$ to $y$ in the time interval $[0,t]$, and 
with the additional assumption that the real-valued coefficients satisfy $c_{xy}=-c_{yx}$. This latter assumption  ensures that currents  are antisymmetric under time reversal.  If the coefficients $c_{xy} \neq -c_{yx}$, then we speak of a general time-additive observable that is not a fluctuating current. To derive  the main  results of this paper we use that currents are   antisymmetric under time-reversal,   and therefore the results of this paper do not directly extend to general time-additive observables.

We investigate in this paper the first-passage problem (\ref{eq:T}) for  fluctuating currents $J$ of the form (\ref{eq:J}).  To fully define $T$, we also need to specify that  $T=\infty$ if $J(t)\in (-\ell_-,\ell_+)$ for all values of $t\geq 0$. According to the D\'{e}but Theorem, see 76.1 in Chapter II of Ref.~\cite{williams}, first-passage times are examples of stopping times,  and therefore we can use Doob's optional stopping theorem \cite{williams1991probability,liptser2001statistics,roldan2022martingales}.

We consider currents with $\langle J(t)\rangle \neq 0$,  as for such currents the first-passage times $T$ at the positive and negative thresholds satisfy a  large deviation principle with speeds $\ell_+$ and $\ell_-$, respectively (see Introduction).   For  such currents,
 without loss of generality, we  can assume that $\langle J(t)\rangle>0$.   
The average rate of the current is  defined as 
\begin{equation}
\overline{j} = \lim_{t\rightarrow \infty}t^{-1}\langle J(t)\rangle. \label{eq:jbar}
\end{equation}
In discrete time the units of time  are set so that the time interval between two steps equals one    (otherwise we need to divide the right-hand side by the time elapsed between two discrete time units).

The scaled cumulants of $J(t)$ can be obtained from the derivatives of the 
 scaled cumulant generating function 
\begin{equation}
\lambda_J(a) = \lim_{t\rightarrow \infty}\frac{1}{t}\ln \langle e^{-a J(t)}\rangle \label{eq:lambdaJ}
\end{equation}
at $a=0$.  For example, $\overline{j} = -\lambda'_J(0)$.  For continuous time Markov chains,  the scaled cumulant generating function $\lambda_J$ equals the Perron root of the tilted matrix~\cite{touchette2009large}
\begin{align}
        \tilde{\mathbf{q}}_{xy} (a) 
        = \left\{\begin{array}{ccc} \exp(-ac_{xy})\mathbf{q}_{xy} && x\neq y,\\ \mathbf{q}_{xx}  &&  x=y.\end{array}\right.  \label{eq:qtilde}
\end{align}
In discrete time, $\lambda_J(a)$  equals the natural logarithm of the Perron root of  $\tilde{\mathbf{q}}(a)$.

\subsection{Entropy production}
The rate of dissipation or entropy production of a Markov jump process is given by~\cite{schnakenberg} 
\begin{equation}
\dot{s} = \sum_{x\in \mathcal{X}}\sum_{y\in \mathcal{X}\setminus \left\{x\right\}} p_{\rm ss}(x)\mathbf{q}_{xy} \ln \frac{\mathbf{q}_{xy}}{\mathbf{q}_{yx}}.\label{eq:sdotSchn}
\end{equation} 
The formula (\ref{eq:sdotSchn}) relies on the physical assumption that the  process $X$ satisfies local detailed balance~\cite{maes2021local}, i.e., the environment is in thermal equilibrium at a temperature $\mathsf{T}_{\rm env}$,  and that $X$ has even parity under time reversal.   As with $\overline{j}$, in discrete time the  unit of time in between two time steps is  set equal to one.

The fluctuating entropy production quantifies  the amount of dissipation along an observed trajectory~\cite{Seki,seifert2012stochastic,peliti2021stochastic, seifert}.  
For stationary Markov jump processes, i.e., when the initial distribution equals the stationary distribution, the fluctuating entropy production is defined by
\begin{equation}
S(t) =  \sum_{x\in \mathcal{X}}\sum_{y\in \mathcal{X}\setminus \left\{x\right\}} \ln \frac{p_{\rm ss}(x)\mathbf{q}_{xy}}{p_{\rm ss}(y)\mathbf{q}_{yx}} N^{xy}(t).  \label{eq:StDef}
\end{equation}
Notice that $S(t)$ is a fluctuating current of the form (\ref{eq:J}), with $c_{xy} = \ln (p_{\rm ss}(x)\mathbf{q}_{xy})/(p_{\rm ss}(y)\mathbf{q}_{yx})$ the entropy produced when the process jumps from $x$ to $y$.   Taking the average we find 
\begin{equation}
\dot{s}= t^{-1}\langle S(t)\rangle. 
\end{equation}
The fluctuating entropy production $S(t)$ satisfies the property 
\begin{equation}
\langle f(X(t))\rangle^\dagger = \langle  f(X(t))e^{-S(t)} \rangle,  \label{eq:Radon}
\end{equation}
where $\langle \cdot \rangle^\dagger$ are averages in the time-reversed Markov chain.    The time-reversed Markov chain has a law $\mathbb{P}^\dagger$ that is defined by the initial condition $p_{\rm ss}(x)$ and the time-reversed transition rate matrix 
\begin{equation}
\mathbf{q}^\dagger = \mathbf{p_{\rm ss}^{-1}} \mathbf{q^T} \mathbf{p_{\rm ss}}, \label{eq:qTimeRev}
\end{equation} 
where $\mathbf{q^T}$ is the transpose of the rate matrix $\mathbf{q}$, $\mathbf{p_{\rm ss}}$ is a diagonal matrix with diagonal entries given by  $p_{\rm ss}(x)$,  and $\mathbf{p_{\rm ss}^{-1}}$ is the matrix inverse of $\mathbf{p_{\rm ss}}$.

\subsection{Tilting and time-reversal}\label{sec:tiltingtimereversal}
In this paper we use repeatedly the following equality 
\begin{equation}
    \mathbf{\tilde{q}^\dagger}(a) = \mathbf{p}_{\rm ss}^{-1} \tilde{\mathbf{q}}^T(-a)\mathbf{p}_{\rm ss},\label{eq:qtildedaggerMT}
\end{equation}
relating the tilted matrix of the time-reversed process, $\mathbf{\tilde{q}^\dagger}$, with the tilted matrix of the forward process $\tilde{\mathbf{q}}$.   Here, the tilted matrix $\mathbf{\tilde{q}^\dagger}$ is defined as in Eq.~(\ref{eq:qtilde}), but with the entries of $\mathbf{q}$ replaced by those of $\mathbf{q}^\dagger$.

A direct consequence of (\ref{eq:qtildedaggerMT}) is that 
\begin{equation}
\lambda_J(a) = \lambda^\dagger_J(-a),   \label{eq:scaledTR}
\end{equation}
where $\lambda^\dagger_J(-a)$ is the scaled cumulant generating function of the time-reversed process, i.e., 
\begin{equation}
\lambda^\dagger_J(a) = \lim_{t\rightarrow \infty}\frac{1}{t}\langle  e^{-aJ(t)}\rangle^\dagger.
\end{equation}
Indeed,  due to (\ref{eq:qtildedaggerMT})  both matrices share the same eigenvalues, and $\lambda_J(a)$ and $\lambda^\dagger_J(-a)$ are determined by the Perron roots of $\tilde{\mathbf{q}}(a)$ and $\tilde{\mathbf{q}}^\dagger(-a)$, respectively.

The Eqs.~(\ref{eq:qtildedaggerMT}) follows readily from the definitions (\ref{eq:qtilde}) and (\ref{eq:qTimeRev})  and the fact that  for fluctuating currents   $c_{xy} = -c_{yx}$ holds for all $x,y\in \mathcal{X}$.   Therefore, the property (\ref{eq:scaledTR}) does not apply to time-additive observables  that are not  fluctuating currents.   As we use the   property (\ref{eq:scaledTR}) to derive the main results of this paper, the results of this paper apply to fluctuating currents.

\subsection{Notation}

We briefly explain the notation used in this paper.  We use upper case roman letters for random variables.  This includes stochastic processes, e.g., $X(t)$ and $J(t)$, random trajectories, e.g., $X_0^T$, or  random times, e.g., $T$.   We use lower case roman letters to indicate their deterministic counterparts, e.g., $x(t)$, $j(t)$, $x^t_0$, and $t$.    We use   $\overline{j}$ for the average rate of the current. Probability mass functions and probability distributions are indicated by  $p$ with a subscript  specifying the distribution being referred to.   We use bold font for matrices, e.g., 
$\mathbf{q}$ and $\mathbf{p_{\rm ss}}$.   We use angular brackets $\langle\cdot\rangle$ to indicate  averages over different realisations of the process $X$, and    expectations conditioned on termination at  $\ell_+$ or $-\ell_-$ are indicated by  $+$ and $-$.   Probabilities and probability measures are denoted by $\mathbb{P}(\cdot)$.   We use the  $\dagger$ symbol  for time reversal, and the $\hat{}$ symbol for the Doob transform of a  tilted Markov chain, the latter which is denoted by     $\tilde{}$.  We use the symbol $\mathcal{I}$ to indicate the  rate function of a  large deviation principle, and we use $\lambda$ or  $m_\pm$  to  indicate the  corresponding scaled cumulant generating function, depending on whether the random variables is a fluctuating current or a first-passage time conditioned on termination at one of the thresholds,  respectively.

\section{Trade-off relation between speed, accuracy, and dissipation}\label{sec:firstbound}
We derive the inequality (\ref{eq:unc}) that expresses a trade-off between speed, accuracy, and dissipation, which are  quantified by the mean first-passage time $\langle T \rangle$, the splitting probability $p_-$, and the dissipation rate $\dot{s}$, respectively. The derivation presented here  applies to both continuous-time and discrete-time Markov chains, and it thus more general than the  derivations for continuous-time Markov chains  that appeared  previously in the literature~\cite{neri2022universal,raghu2024effective}.

We organise the derivation  of (\ref{eq:unc}) in three parts.  In Sec.~\ref{sec:coarsegrain} we focus on the quantity $\langle S(T)\rangle$, which is the average of the entropy production evaluated at the first-passage time $T$~\cite{neri2019integral,neri2020second,gonz}.   We show that $\langle S(T)\rangle$  decreases upon coarse-graining  the trajectories $X^T_0$.   With a proper choice of the coarse-grained variable,   we obtain a lower bound for  $\langle S(T)\rangle$ in terms of the splitting probabilities of $T$ in the forward and time-reversed dynamics.    In the following Sec.~\ref{sec:threshold}, we take the large thresholds limit and we derive in this limit an asymptotic Wald equality that relates $\langle S(T)\rangle$  to the rate of dissipation, $\dot{s}$, and the mean first-passage time, $\langle T\rangle$.    Lastly, in  Sec.~\ref{sec:mart}, we  show that in the limit of large thresholds the   splitting probabilities in the time-reversed dynamics are related to  those in the forward dynamics, which completes the derivation.     

Note that the restriction to continuous time in previous derivations of Eq.~(\ref{eq:unc}) stems from the reliance on the inequality $\lambda_J(a)\geq a\overline{j}(-1+a\overline{j}/\dot{s})$~\cite{pietzonka,gingrich2017fundamenta}, which does not hold in discrete time~\cite{proesmans2017discrete}. In contrast, here we get the  inequality in (\ref{eq:unc}) by coarse-graining  $\langle S(T) \rangle$, which applies in both continuous and discrete time

\subsection{Inequality for $\langle S(T)\rangle$ in terms of the splitting probabilities of $T$}\label{sec:coarsegrain}
We show that  the average entropy production  $\langle S(T)\rangle$ evaluated at  a first passage time $T$  is bounded from below   by 
\begin{equation}
\langle S(T)\rangle \geq p_+ \ln \frac{p_+}{p^\dagger_+} + p_- \ln \frac{p_-}{p^\dagger_-}, \label{eq:ST}
\end{equation} 
where 
\begin{equation}
p_+ = \mathbb{P}\left(J(T)\geq \ell_+\right) \quad {\rm and} \quad   p_- = \mathbb{P}\left(J(T)\leq -\ell_- \right) 
\end{equation} 
are the splitting probabilities of $T$  in the forward dynamics, and  where 
\begin{equation}
p^\dagger_+ = \mathbb{P}^\dagger\left(J(T)\geq \ell_+\right) \quad  {\rm and} \quad  p^\dagger_- = \mathbb{P}^\dagger\left(J(T)\leq -\ell_- \right) , \label{eq:splittingTR}
\end{equation}
are the splitting probabilities of $T$ in the  time-reversed dynamics.    

To derive (\ref{eq:ST}),    we first show in Sec.~\ref{sec:KLDLike} that $\langle S(T)\rangle$ is  a  Kullback-Leibler divergence between two probability distributions defined on the set of realisations of the trajectories $X^T_0$.    Then in Sec.~\ref{sec:coarseGrainST}, we obtain (\ref{eq:ST}) through a  coarse-graining scheme applied to $X^T_0$.

 \subsubsection{Expressing $\langle S(T)\rangle$ as a Kullback-Leibler divergence. }\label{sec:KLDLike}
The definition (\ref{eq:Radon})  of $S(t)$ implies that $\exp(-S(t))$ is the Radon-Nikodym derivative process~\cite{seifert2012stochastic, neri2017statistics, yang}, 
\begin{equation}
e^{-S(t)} = \frac{p^\dagger(X^t_0)}{p(X^t_0)},
\end{equation}
where $p(X^t_0)$  
and $p^\dagger(X^t_0)$ are the stationary probability distributions associated with the trajectories  $X^t_0$ in the forward and time-reversed dynamics.  
For stationary Markov chains in discrete time, $p(X^t_0)$   is the probability mass function defined in Eq.~(\ref{eq:pPathDiscrete})  with $p_{\rm init} = p_{\rm ss}$,  
and $p^\dagger(x^t_0)$ is the corresponding path probability mass function of the time-reversed Markov chain with transition probability $\mathbf{q}^\dagger$ and initial state $p_{\rm ss}$.     For Markov chains in continuous time, $p(X^t_0)$   is as defined in Eq.~(\ref{eq:PathContinuous})  with $p_{\rm init} = p_{\rm ss}$, and $p^\dagger$ is the corresponding path probability distribution of the time-reversed Markov chain with transition rate matrix $\mathbf{q}^\dagger$ and initial state $p_{\rm init} = p_{\rm ss}$.

The quantity $\langle S(T)\rangle$  is the average of $S(t)$ evaluated at the first-passage time $t=T(X^{\infty}_0)$.  In discrete time, it follows from the definition of $\langle \cdot\rangle$ that the average  of $S(T)$ over all trajectories can be expressed as 
\begin{eqnarray}
\langle S(T)\rangle &=& \sum_{x^{\infty}_0\in \Omega}  p(x^{\infty}_0) \ln \frac{p(x_0^{T(x^{\infty}_0)})}{p^\dagger(x_0^{T(x^{\infty}_0)})}, \label{eq:STDef}
\end{eqnarray} 
where $p(x_0^{T(x^{\infty}_0)})$ is the path probability $p(x^t_0)$ for $t=T$ (and similarly for $p^\dagger(x_0^{T(x^{\infty}_0)})$). 
Note that $p(x^{\infty}_0)$ is a probability distribution over all trajectories in $\Omega$, but  $S(T)$ only depends on the part of the trajectory before the first-passage time.  Therefore, we can ``marginalise" the distribution $p(x^{\infty}_0)$ in the right-hand side of Eq.~(\ref{eq:STDef}) in the following way 
\begin{eqnarray}
\langle S(T)\rangle &=& \sum_{x^{\infty}_0\in \Omega} \sum^{\infty}_{t=0}\delta(t;T(x^\infty_0)) p(x^{\infty}_0) \ln \frac{p(x_0^t)}{p^\dagger(x_0^t)}\nonumber\\ 
&=& \sum^{\infty}_{t=0}\sum_{x_0^t\in \mathcal{X}^{t+1}} \delta(t;T(x^t_0)) p(x_0^t) \ln \frac{p(x_0^t)}{p^\dagger(x_0^t)}, \label{eq:expression}
\end{eqnarray} 
where $\delta(\cdot;\cdot)$ is a Kronecker delta function.  The first  equality introduces a Kronecker delta function to replace $T$ by the summation index $t$. The last equality uses the stopping time property $T(x^{\infty}_0) = T(x^T_0)$ of the first-passage time~\cite{williams1991probability, roldan2022martingales}, and then marginalises the distribution $p(x^{\infty}_0)$ by summing over all variables $x(t')$ with $t'>t$.  Note that we have used the notation $T(x^t_0)  = T(x^{\infty}_0)$ if $t\geq  T(x^{\infty}_0)$, and $T(x^t_0)=\infty$ if   $t< T(x^{\infty}_0)$.     We can express  the right-hand side of (\ref{eq:expression})  into the more compact form 
\begin{eqnarray}
\langle  S(T) \rangle  &=& \sum^{\infty}_{t=0}\sum_{x_0^t\in \mathcal{X}^{t+1}}' p(x_0^t) \ln \frac{p(x_0^t)}{p^\dagger(x_0^t)},\label{eq:STconditional} 
\end{eqnarray} 
where the prime indicates that we only sum over the trajectories for which $T(x^t_0) = t$.  The average $\langle  S(T) \rangle$ can thus be seen as an average over the ensemble of ``stopped" trajectories  terminating at their respective first passage times $T$.

The Eq.~(\ref{eq:STconditional})  makes it clear that $\langle S(T)\rangle$ is the Kullback-Leibler divergence of the probability distribution $p$ from $p^\dagger$.  Here, both probability distribution are defined on the set of realisations of the ``stopped" trajectories $X^T_0$. Note that these probability distributions are properly normalised as $\sum^{\infty}_{t=0}\sum_{x_0^t\in \mathcal{X}^{t+1}}' p(x_0^t) = 1$.    Hence, both    $\langle S(T)\rangle$ and $\langle S(t)\rangle$ are Kullback-Leibler divergences between  probability distributions  of trajectories related to $\mathbb{P}$ and $\mathbb{P}^\dagger$, but the distributions are supported on different sets of trajectories.

The  expression (\ref{eq:STconditional}) generalises to  continuous time
\begin{equation}
\langle  S(T) \rangle  = \int^{\infty}_0 dt   \sum'_{x_0^t \in \mathcal{X}^{t+1}} p(x_0^t)  \ln \frac{p(x_0^t)}{p^\dagger(x_0^t)}. \label{eq:KLD}
\end{equation} 
Here, $\sum'_{x_{[0,t]}\in \mathcal{X}^{t+1}}$  denotes a generalised sum, understood as a combination of summations and integrals over all trajectories satisfying $T(X^t_0) = t$.   This summation can be expressed as an integral over all jump times with  the condition that the last jump time equals $t$, i.e., $t(n)=t$, and a sum over all discrete-time trajectories $y^{n}_0$ with the constraint that  the corresponding first-passage problem in discrete time satisfies $T(y^n_0) = n$.
For simplicity, in what follows we perform the calculations in discrete time, using the formula \eqref{eq:STconditional} for $\langle S(T)\rangle$ in discrete time.   However, the derivations also hold in continuous time and can be obtained by replacing the summation over time with an integral in the steps that follow.
 
\subsubsection{Coarse-graining of $\langle S(T)\rangle$.}\label{sec:coarseGrainST} 
The average entropy production $\langle S(t)\rangle$ decreases under coarse-graining, see, e.g., Refs.~\cite{coarseGraining,coarseGraining2, harunari_what_2022,partial, blom2024milestoning,ertel2024estimator,fritz2025entropy}.   This behaviour under coarse-graining follows from the fact that $\langle S(t)\rangle$ is a Kullback-Leibler divergence.   
As shown in the previous section,  also    $\langle S(T)\rangle$ is  a Kullback-Leibler divergence, and therefore we can also coarse-grain  this quantity.   However,   while $\langle S(t)\rangle$ involves probability distributions defined on the set of realisations of $X^t_0$, the 
the  Kullback-Leibler divergence $\langle S(T)\rangle$  involves probability distributions defined on the set of realisations of the trajectories $X^T_0$.    Consequently, when coarse-graining $\langle S(T)\rangle$, we need to use coarse
observables that are functionals of $X^T_0$.  In what follows, we implement this procedure for a specific observable in discrete-time Markov chains that yields (\ref{eq:ST}), and we refer to~\ref{app:conv} for a  derivation that applies to arbitrary observables in discrete-time and continuous-time Markov chains. 

We consider the coarse observable
\begin{equation}
D = {\rm sign}\left(J(T)\right), 
\end{equation}
with   $D=0$ if $T=\infty$.    Considering that $\mathbb{P}(D=0)=0$, we 
 can express the formula (\ref{eq:STconditional}) for $\langle S(T)\rangle$ as 
\begin{eqnarray}
 \langle  S(T) \rangle  = -\sum_{d\in \left\{ \pm \right\}}  p_d \sum^{\infty}_{t=0}\sum'_{x^t_0\in \mathcal{X}^{t+1}}
\delta(d;D(x^t_0))\: p(x^t_0|D=d)  \: \ln \frac{p^\dagger(x^t_0)}{  p(x^t_0)}, \label{eq:STpD}
\end{eqnarray}
where  $p(x^t_0|D=d) =  p(x^t_0)/p_d$, and we recall that $p_d$ are the splitting probabilities of $T$ (with $d=+$ or $d=-$).    Note that in  Eq.~(\ref{eq:STpD})  we have used that $D$ is a functional of the trajectories $X^T_0$ terminating at the stopping time $T$, as the sum in the right-hand side of (\ref{eq:STpD}) runs over all realisations of these ``stopped" trajectories $X^T_0$.  

Applying Jensen's inequality~\cite{Jensen1906, Durrett2019-os} to the expectation value of a convex function, which here is  $-\ln (\cdot)$, we obtain 
\begin{eqnarray}
  \langle  S(T) \rangle  &\geq&  -\sum_{d\in \left\{ \pm 1\right\}}  p_d  \ln \left(\sum^{\infty}_{t=0} \sum'_{x^t_0\in \mathcal{X}^{t+1}} \delta(d;D(x^t_0)) p(x^t_0|D=d)\frac{
p^\dagger(x^t_0)}{ p(x^t_0)}\right) \nonumber\\ 
&=& -\sum_{d\in \left\{ \pm 1\right\}}p_d \ln \frac{p^\dagger_d}{p_d}, \label{eq:SIntermediate}
\end{eqnarray}
which we recognise as the inequality (\ref{eq:ST}) with $p^\dagger_d$  the splitting probabilities of $T$ in the time-reversed dynamics.   Note that the expectation value that we used for  Jensen's inequality is with respect to  a probability measure defined on the set of realisations of the trajectories $X^T_0$ and with probability mass function $p(x^t_0|D=d)$.

\subsection{Limit of large thresholds}\label{sec:threshold}
We show that in the limit of large thresholds, 
\begin{equation}
\dot{s}  \geq  \frac{|\ln p^\dagger_+|}{\langle T\rangle}(1+o_{\ell_{\rm min}}(1)),  \label{eq:asymptotic}
\end{equation} 
where $o_{\ell_{\rm  min}}(1)$ is a function that converges to zero when $\ell_{\rm min} = {\rm min}\left\{\ell_+,\ell_-\right\}$ diverges.   Note that this inequality appeared before in Refs.~\cite{decision1,neri2022universal} by applying results from  the theory of sequential hypothesis testing to stochastic thermodynamics~\cite{tartakovsky2014sequential}.

As shown in \ref{app:B1}, in the limit of large thresholds
\begin{equation}
p_-  = \exp\left(-a^\ast \ell_- [1+ o_{\ell_{\rm min}}(1)] \right) \label{eq:pM}
\end{equation}
with $a^\ast>0$ the effective affinity, i.e, the nonzero value of $a$ for which 
the scaled cumulant generating function $\lambda_J(a)=0$, i.e., 
\begin{equation}
\lambda_J(a^\ast) = 0.
\end{equation}
  As the average value of $J(t)$ in the time-reversed process is negative,  
\begin{equation}
p^\dagger_+  = \exp\left(a^{\dagger, \ast} \ell_+ [1+ o_{\ell_{\rm min}}(1)] \right), \label{eq:pPDagger}
\end{equation}
where  $a^{\dagger, \ast}<0$ is  the effective affinity of $J$ in the time-reversed dynamics.  In other words,   $a^{\dagger, \ast}$ is the nonzero value of $a$ for which $\lambda^\dagger_J(a)=0$.  Using the scaling laws (\ref{eq:pM}) and (\ref{eq:pPDagger}), along with $p_-+p_+ = p^\dagger_- + p^\dagger_+ = 1$, we find that in the limit of large thresholds (\ref{eq:ST}) simplifies into
\begin{equation}
\langle S(T)\rangle  \geq  |\ln  p^\dagger_+| (1+o_{\ell_{\rm min}}(1)).    \label{eq:STAv2aM}
\end{equation}

The left-hand side of (\ref{eq:STAv2aM}) can be approximated using an asymptotic version of  Wald's equation~\cite{Wald2,Wald1}
\begin{equation}
\langle S(T)\rangle = \dot{s}\langle T\rangle (1+o_{\ell_{\rm min}}(1)).\label{eq:STAvM}
\end{equation}
In ~\ref{app:E}  we derive (\ref{eq:STAvM}) for first-passage times with two thresholds by using the ergodicity of the process $X(t)$.
Combining (\ref{eq:STAvM}) with (\ref{eq:STAv2aM}) yields (\ref{eq:asymptotic}).
\subsection{Splitting probabilities of fluctuating currents in the time-reversed process}\label{sec:mart}
We demonstrate here that due to the anti-symmetry of fluctuating currents under time reversal,
\begin{equation}
|\ln p^\dagger_+| = \frac{\ell_+}{\ell_-} |\ln p_-|(1+o_{\ell_{\rm min}}(1)),  \label{eq:pDagger}
\end{equation} 
which together with (\ref{eq:asymptotic})  completes the derivation of the inequality (\ref{eq:unc}).      Note that the equality  (\ref{eq:pDagger}) was  conjectured in Section 5 of  Ref.~\cite{neri2022universal}   based on physical intuition, but a convincing derivation was missing.        In   the present paper we derive  (\ref{eq:pDagger}) by using recent results on martingales~\cite{raghu2024effective, neri2025martingale}. 

As shown in \ref{app:B1},  the exponential decay constant $a^\ast$  of $p_-$ is the nonzero root of the equation
\begin{equation}
\lambda_J(a^\ast) = 0, \label{eq:EADef}
\end{equation} 
and analogously we find 
\begin{equation}
\lambda^\dagger_J(a^{\dagger,\ast}) = 0
\end{equation} 
for the exponential decay constant of the time-reversed splitting probability $p^\dagger_+$. As $J$ is antisymmetric under time reversal, we can use the Eq.~(\ref{eq:scaledTR})  that relates  $\lambda_J$ with $\lambda^\dagger_J$.   Using (\ref{eq:scaledTR}) we readily recover from the above two equations that the effective affinity changes sign under time reversal, i.e.,
\begin{equation}
a^{\dagger,\ast}= -a^\ast. \label{eq:EATR}
\end{equation} 
Using (\ref{eq:EATR}) in  Eqs.~(\ref{eq:pM}) and (\ref{eq:pPDagger}), we obtain   (\ref{eq:pDagger}).

\section{Refined inequality involving the statistics of  $T$ at the negative threshold} \label{sec:PT} 
In the previous section we have derived a bound on the rate of dissipation $\dot{s}$ by coarse-graining $\langle S(T)\rangle$ with the observable $D(X^T_0)$.  This approach works more generally for observables that are functionals of $X^T_0$.  Hence, by refining the observables that are being measured, we can obtain better bounds on the rate of dissipation.    In the present section, we coarse-grain $\langle S(T)\rangle$ by using the pair $(D,T)$ of coarse observables.   In this way, we will obtain the bounds Eqs.~(\ref{eq:totalbound}) and (\ref{eq:newboundIJform}).

As in the previous section, we organise the derivation of (\ref{eq:totalbound})  in three  parts: in Sec.~\ref{sec:41} we  bound $\langle S(T)\rangle$ from below through the above mentioned coarse-graining procedure; in Sec.~\ref{sec:42}, we take the limit of large thresholds; and  in Sec.~\ref{sec:43} we relate the first-passage time distributions in the time-reversed process to those in the forward process.  In Sec.~\ref{sec:improvedBound} we derive the Eq.~(\ref{eq:newboundIJform}) that expresses the bound (\ref{eq:totalbound}) in terms of the fluctuations of $J(t)$ at fixed time $t$.

\subsection{Coarse-graining of the average entropy production $\langle S(T)\rangle$}\label{sec:41}
We show that 
\begin{equation}
{\langle S(T)\rangle \geq   p_+ \ln \frac{p_+}{p^\dagger_+} +  p_+ \ln \frac{p_+}{p^\dagger_+}  } + p_+ \sum^{\infty}_{t=0} p_T(t|+) \ln \frac{p_T(t|+)}{p^\dagger_T(t|+)} + p_- \sum^{\infty}_{t=0}   p_T(t|-) \ln \frac{p_T(t|-)}{p^\dagger_T(t|-)}, \label{eq:refinedIneq}
 \end{equation}
 where $p^\dagger_T(t|+)$ and $p^\dagger_T(t|-)$ are the probability mass functions of $T$ in the time-reversed dynamics conditioned on $J(T)\geq \ell_+$ and $J(T)\leq -\ell_-$, respectively.  In continuous time, we should replace the sums over $t$ by integrals and    $p_T(t|\pm)$ becomes a probability distribution. 

To derive (\ref{eq:refinedIneq})  we use a similar argument as in Sec.~\ref{sec:coarseGrainST}.      The main distinction is in the application of Jensen's inequality, which now will be used for expectation values with respect to $p(x^t_0|D=d,T=t)$.   To this purpose, we express (\ref{eq:STpD})  as 
\begin{eqnarray}
  \langle  S(T) \rangle  = -\sum_{d\in \left\{ \pm \right\}}  p_d \sum^{\infty}_{t=0}p_T(t|D=d)\sum'_{x^t_0\in \mathcal{X}^{t+1}}
\delta(d;D(x^t_0))\: p(x^t_0|D=d,T=t)  \: \ln \frac{p^\dagger(x^t_0)}{  p(x^t_0)}.\nonumber\\ \label{eq:STpDv2}
\end{eqnarray}
Applying Jensen's inequality to the expectation of the convex function $-\ln(\cdot)$, as in Eq.~(\ref{eq:SIntermediate}), we now obtain (\ref{eq:refinedIneq}).

 \subsection{Limit of large thresholds}\label{sec:42}
We show that in the limit of large thresholds, 
\begin{equation}
\dot{s}\geq  \frac{\ell_+}{\langle T\rangle} \left(\frac{|\ln p^\dagger_+|}{\ell_-} +   \mathcal{I}^\dagger_+(1/\overline{j}) \right)(1+o_{\ell_{\rm min}}(1)). \label{eq:boundO}
\end{equation} 
 
 In the limit of large thresholds using Eqs.~(\ref{eq:pM}) and (\ref{eq:pPDagger}),  we obtain 
 \begin{equation}
\langle S(T)\rangle \geq     \left(|\ln p^\dagger_+|  + p_+ \sum^{\infty}_{t=0} p_T(t|+) \ln \frac{p_T(t|+)}{p^\dagger_T(t|+)} \right)(1+o_{\ell_{\rm min}}(1)). \label{eq:58}
 \end{equation}
For  $\overline{j}\neq 0$ and $\sigma^2_J>0$, the distributions $p_T(t|+)$ and $p^\dagger_T(t|+)$ satisfy a large deviation principle with speed $\ell_+$ and rate functions~\cite{gingrich2017fundamenta} 
\begin{equation}
 \mathcal{I}_+(\tau) = \lim_{\ell_+\rightarrow \infty} 
 \frac{|\ln  p_{T}(\ell_+\tau|+) |}{\ell_+} \quad {\rm and} \quad   \mathcal{I}^\dagger_+(\tau) = \lim_{\ell_+\rightarrow \infty} 
 \frac{|\ln  p^\dagger_{T}(\ell_+\tau|+) |}{\ell_+} . \label{eq:largedev}
 \end{equation} 
 By using in Eq.~(\ref{eq:58})  the large deviation principles (\ref{eq:largedev}) together with 
  the asymptotic version of Wald's equation, Eq.~(\ref{eq:STAvM}), we find that 
\begin{equation}
\dot{s} \geq   \frac{1}{\langle T\rangle} \left(\frac{\ell_+}{\ell_-}|\ln p_-| + \ell_+  \int^{\infty}_{0} {\rm d}\tau  \: p_{T}(\ell_+\tau|+)  \: [\mathcal{I}^\dagger_+(\tau)-\mathcal{I}_+(\tau)] \right)(1+o_{\ell_{\rm min}}(1)).
 \end{equation}
 Using that  in the limit of large thresholds, $\ell_+\gg1$, it holds that  $p_{T}(\ell_+\tau|+)\approx \delta(\tau-\langle T|+\rangle/\ell_+)$, and using the following asymptotic version of Wald's equation (see Appendix C in \cite{raghu2024effective})
 \begin{equation}
 \frac{1}{\overline{j}}= \lim_{\ell_+\rightarrow \infty}\frac{\langle T|+\rangle}{\ell_+} = \lim_{\ell_+\rightarrow \infty}\frac{\langle T\rangle}{\ell_+},
 \label{eq:asympWald}
 \end{equation}
 we  obtain Eq.~(\ref{eq:boundO}).

  \subsection{Time-reversal of the rate functions of $T$}\label{sec:43}
  We show  that for large enough thresholds the statistics of $T$ at the negative threshold in the forward dynamics equal the statistics of $T$ at the positive threshold in the time-reversed dynamics, i.e.,
 \begin{equation}
  \mathcal{I}_-(\tau) = \mathcal{I}^\dagger_+(\tau),  \label{eq:ITR}
 \end{equation}  
 Equation (\ref{eq:ITR}) together with (\ref{eq:boundO})  completes the derivation of the inequality (\ref{eq:totalbound}). 
 Just as was the case for the derivation of the equality (\ref{eq:pDagger}), the property (\ref{eq:ITR}) relies on  the anti-symmetry of fluctuating currents under time-reversal.

 To derive (\ref{eq:ITR}), 
we use  the fact that the scaled cumulant generating function $m_-$ [as defined in Eq.~(\ref{eq:mDef})] is the functional inverse of $\lambda_J$~\cite{gingrich2017fundamenta, neri2025martingale}, viz., 
 \begin{equation}
\lambda_J(a^\ast-m_-(\mu)) = -\mu, \label{eq:11}
\end{equation}
where $a^\ast$ is the effective affinity.   This  result implies that the large deviation properties of $T$ are determined by those of $J$ (see~\ref{app:B2} for a derivation of (\ref{eq:11}) using martingales). 
Analogously,  the scaled cumulant generating 
\begin{equation}
m^\dagger_+(\mu) =\lim_{\ell_+\rightarrow \infty} \frac{1}{\ell_+}\langle e^{\mu T}\rangle^\dagger_+ 
\end{equation}
is the functional inverse of $\lambda^\dagger_J$, i.e., 
 \begin{equation}
\lambda^\dagger_J(a^{\dagger,\ast}+m^\dagger_+(\mu)) = -\mu,  \label{eq:12}
\end{equation} 
where $a^{\dagger,\ast}$ is the effective affinity of $J$ in the time-reversed dynamics.  We recall that  $a^{\dagger,\ast} = -a^\ast$ [see Eq.~(\ref{eq:EATR})].
Using  (\ref{eq:11}) and (\ref{eq:12}) in  Eq.~(\ref{eq:scaledTR}), we find
$a^\ast - m_-(\mu) = -a^{\dagger,\ast} -m^\dagger_+(\mu)$,
and thus from (\ref{eq:EATR}) we recover 
\begin{equation}
m_-(\mu)  = m^\dagger_+(\mu).   \label{eq:mScaled}
\end{equation}
Since the large deviation rate functions $\mathcal{I}_{\pm}(\tau)$ are given by the Legendre transforms
\begin{equation}
 \mathcal{I}_-(\tau) = {\rm max}_{\mu} (\mu\tau-m_-(\mu)) \quad {\rm and} \quad \mathcal{I}^\dagger_+ = {\rm max}_{\mu}(\mu \tau - m^\dagger_+(\mu)),\label{eq:IpmLegendre}
\end{equation}
the symmetry (\ref{eq:mScaled}) implies (\ref{eq:ITR}), which is the equation we were meant to demonstrate. 
 
\subsection{Derivation of Eq.~(\ref{eq:newboundIJform})} \label{sec:improvedBound}
We derive the inequality (\ref{eq:newboundIJform}) from \eqref{eq:totalbound}.   Using the Eq.~\eqref{eq:thermoSplit} for the exponential decay rate of $p_-$ and the asymptotic  version of Wald's equation, (\ref{eq:asympWald}), we may rewrite  \eqref{eq:totalbound} as 
\begin{equation}
    \dot{s}\geq \overline{j} \left(a^\ast + \mathcal{I}_-\left(1/\overline{j}\right)\right)(1+o_{\ell_{\rm  min}}(1))   .\label{eq:refinedboundsimple} 
\end{equation}

Next, we  use  in Eq.~(\ref{eq:refinedboundsimple}) the relation \cite{gingrich2017fundamenta,neri2025martingale}
\begin{equation}
   j \: \mathcal{I}_-(1/j) = \mathcal{I}_J(-j) - a^\ast {j} ,\label{eq:IminusIJ}
\end{equation}
which expresses the rate function   $\mathcal{I}_-$ of the first-passage time $T$ at the negative threshold in terms of the rate function $ \mathcal{I}_J(j) = \lim_{t\to\infty} \vert p_{J/t}(j) \vert/t$ of $J(t)$, yielding
\begin{equation}
\dot{s} \geq \mathcal{I}_J(-\overline{j}).  \label{eq:sdotIJ}
\end{equation}
Since the  rate function $\mathcal{I}_J$ is   the Fenchel-Legendre transform of $\lambda_J$~\cite{touchette2009large}, 
\begin{equation}
    \mathcal{I}_J(j)=\max _{a \in \mathbb{R}}\left(-a j-\lambda_J(a)\right),\label{eq:IJGartnerEllis}
\end{equation}
we can express Eq.~\eqref{eq:sdotIJ} as  Eq.~\eqref{eq:newboundIJform}, which is what we were meant to show. 

\begin{figure}[h!]{
    \centering
    \includegraphics[width=0.5\linewidth]{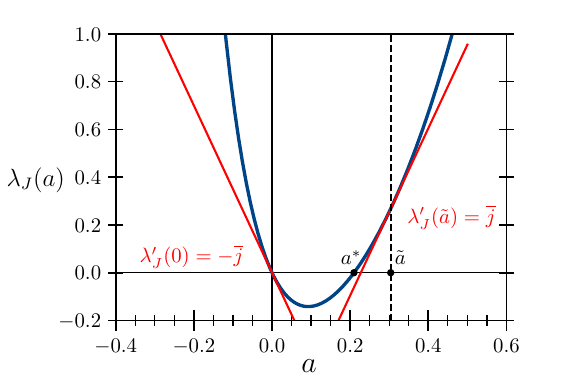}
    \caption{A graphical illustration of the refined bound Eq.~\eqref{eq:newboundIJform} for the rate of dissipation $\dot{s}$.   The scaled cumulant generating function, $\lambda_J(a)$, is plotted as a function of $a$ (blue solid line), and the tangents of $\lambda_J(a)$ at $a=0$ and $a=\tilde{a}$ are indicated (red solid lines).    
    The  right-hand side of Eq.~\eqref{eq:newboundIJform} is $\mathcal{I}_J(-\overline{j})$, which is equal to  the Legendre transform of $\lambda_J(a)$ evaluated at  $a=\tilde{a}$ where $\lambda_J'(\tilde{a}) = \overline{j}$.  Instead, the  right-hand side of the coarser bound  \eqref{eq:sjas} is given by $a^\ast \overline{j}$ with  $\lambda_J(a^\ast)=0$. Note also that $\lambda_J(0)=0$ and $\lambda_J'(0)=-\overline{j}$.  The two bounds are equivalent when $\tilde{a} = a^\ast$, which is equivalent with the symmetry \eqref{eq:weaksymm}.}
    \label{fig:fig2}
}\end{figure}

In Fig.~\ref{fig:fig2}, we illustrate Eq.~\eqref{eq:newboundIJform} by plotting $\lambda_J(a)$ for a generic current $J(t)$ and indicating the quantities $\overline{j}$, $a^\ast$, and $\tilde{a}$ [as defined in (\ref{eq:newboundIJform})].

\section{Symmetries in the first-passage times of fluctuating currents }

As mentioned in the introduction,  it has previously been observed that symmetries of currents are related  to the current optimality~\cite{raghu2024effective}, and thus symmetry is related to  the amount of dissipation that one can infer from that current.       In Sec.~\ref{sec:symmetry} we further clarify  the connection between current symmetry and optimality.  Subsequently, in Sec.~\ref{sec:FT} we derive a  symmetry relation that  can be seen as an extension  of the first-passage time symmetry (\ref{eq:opt})  to generic currents. 

\subsection{Symmetries for optimal currents}\label{sec:symmetry}
Optimal currents are defined as currents for which the equality in Eq.~(\ref{eq:sjas}) is attained~\cite{raghu2024effective}, i.e., 
\begin{equation}
\dot{s} = \overline{j}a^\ast. \label{eq:EA}
\end{equation} 
These currents optimise  the trade-off relation between uncertainty, speed, and dissipation as expressed by Eq.~(\ref{eq:sjas}).   Furthermore, for optimal currents all the  dissipation contained in the process $X$ is captured by estimates based on the effective affinity using Eq.~(\ref{eq:EA}). 

In Ref.~\cite{raghu2024effective}, it was shown that the fluctuating entropy production $S(t)$ is optimal, and that currents belonging to the same cycle equivalence class as $S(t)$ are also optimal. These currents also satisfy the Gallavotti-Cohen type fluctuation symmetry~(\ref{eq:GCsymLDR}) and the corresponding symmetry (\ref{eq:opt}) for the statistics of first-passage times, which suggests a relation between symmetry and optimality.    However, the Gallavotti-Cohen fluctuation symmetry is not a sufficient condition for optimality, as there exist currents that satisfy this symmetry but are not optimal~\cite{raghu2024effective}.  In addition,  it has not been proven that the Gallavotti-Cohen fluctuation symmetry is a necessary condition.

Here, we show  that optimal currents satisfy the symmetry condition 
\begin{equation}
\lim_{\ell_+\rightarrow \infty}\frac{\langle T\rangle_+}{\ell_+} = \lim_{\ell_-\rightarrow \infty}\frac{\langle T\rangle_-}{\ell_-} ,   \label{eq:symm2}
\end{equation}  
and thus a necessary condition for optimality is that the average speed of reaching the positive threshold equals the average speed of reaching the negative threshold.    Indeed, Eq.~(\ref{eq:symm2}) readily follows from Eq.~(\ref{eq:totalbound}).  The  Eq.~(\ref{eq:totalbound}) together with the definition of optimality implies that $\mathcal{I}_-(1/\overline{j})=0$.  Since  $1/\overline{j} = \lim_{\ell_+\rightarrow \infty} \langle T\rangle_+/\ell_+$, and  since for a large-deviation rate function $\mathcal{I}_-(\tau)=0$ if and only if $\tau=\lim_{\ell_+\rightarrow \infty}\langle T\rangle_-/\ell_-$, the symmetry relation (\ref{eq:symm2}) follows from the necessary condition $\mathcal{I}_-(1/\overline{j})=0$.   Note that  Eq.~(\ref{eq:symm2})  is a necessary and sufficient condition for the equality of the right-hand sides of the bounds \eqref{eq:unc} and \eqref{eq:totalbound}.

Examples of currents that satisfy the symmetry \eqref{eq:symm2} are currents that satisfy the Gallavotti-Cohen fluctuation symmetry (\ref{eq:GCsymLDR}), but they are not the only kind of currents for which \eqref{eq:symm2} holds.   Indeed, for currents that satisfy the Gallavotti-Cohen fluctuation symmetry the rate functions at the positive and negative thresholds are the same, see Eq.~(\ref{eq:opt}), whereas \eqref{eq:symm2}  is the weaker statement that the locations of the minima  of the rate functions are the same.  

We can also express the  symmetry \eqref{eq:symm2}  as a property of $\mathcal{I}_J$ or $\lambda_J(a)$.   Specifically,  the symmetry \eqref{eq:symm2} is  equivalent to 
\begin{equation}
    \mathcal{I}_J(-\overline{j}) - a^\ast \overline{j}= \mathcal{I}_J(\overline{j})=0,\label{eq:symm2IJ}
\end{equation}
which we recognise as Eq.~(\ref{eq:GCsymLDR}) for $j=\overline{j}$.  In terms of $\lambda_J(a)$  we get the condition
\begin{equation}
a^\ast = \tilde{a},
\end{equation}
where  $\lambda_J(a^\ast)=0$ and $\lambda'_J(\tilde{a})=\overline{j}$, 
as illustrated in Fig.~\ref{fig:fig2}.

\begin{figure}[h!]{
\centering
\includegraphics[width=0.48\textwidth]{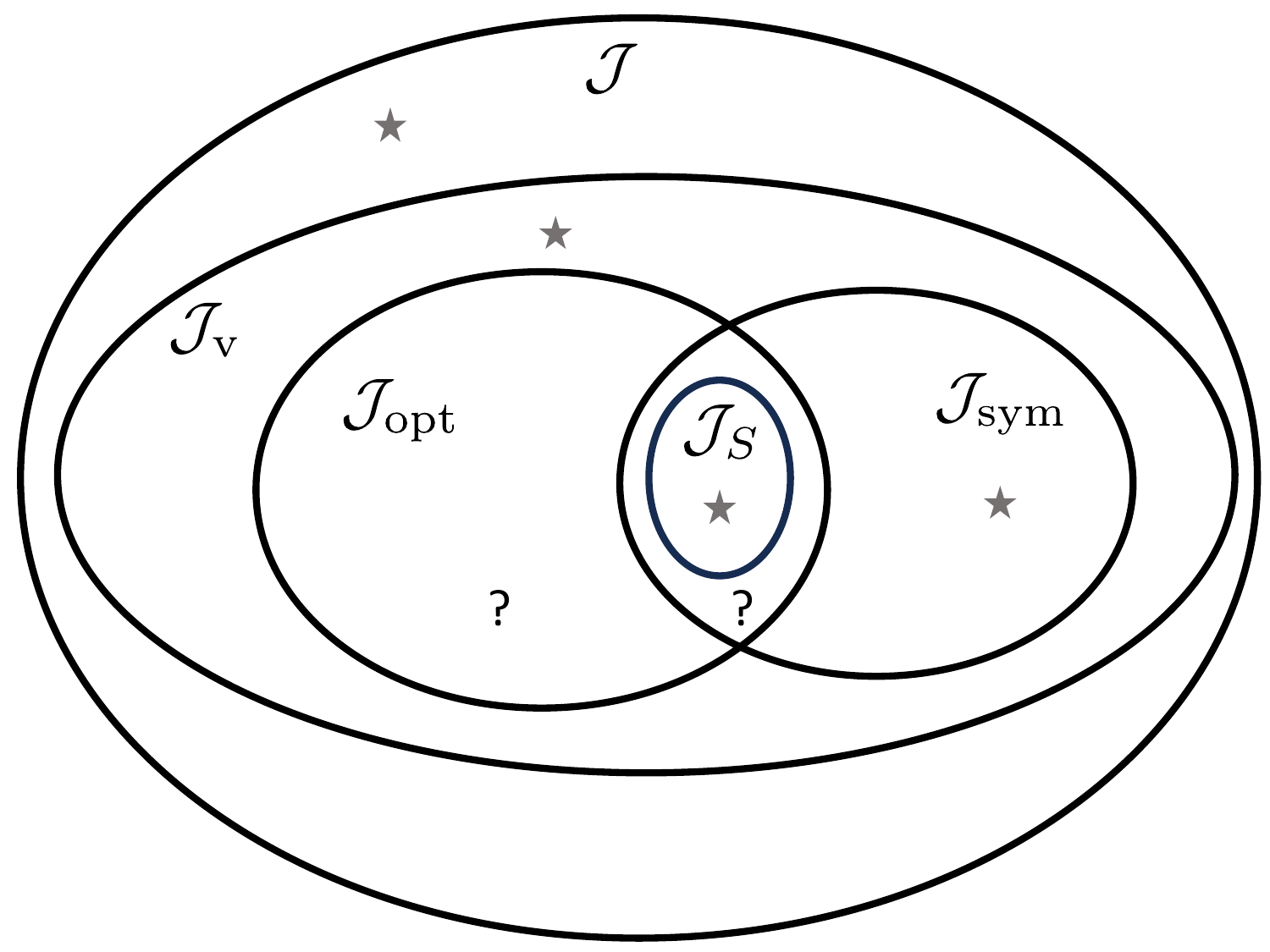}
\caption{Classification of currents according to symmetries and optimality.  Relationships between sets of currents with different properties are shown (see Sec.~\ref{sec:symmetry} for definitions). A star indicates a set for which example currents are known to exist, while a question mark denotes a set whose non-emptiness remains unknown.  }
\label{fig:fig3}   
}\end{figure}

Taken together, there exists an interesting relationship between   current symmetry and optimality.     We summarise this in  Fig.~\ref{fig:fig3} with a Venn diagram that represents the following sets of fluctuating currents:
\begin{itemize}
\item The set $\mathcal{J}$  is the set of all fluctuating currents of $X$ as defined in Eq.~(\ref{eq:J}), namely,
\begin{equation}
\mathcal{J} = \left\{ \sum_{x\in \mathcal{X}}\sum_{y\in \mathcal{X}\setminus \left\{x\right\}}c_{xy}N^{xy}(t) : \forall x,y\in \mathcal{X}^2,  c_{xy}=-c_{yx}\in \mathbb{R}\right\}.
\end{equation}
Note that $\mathcal{J}$ is isomorphic with $\mathbb{R}^{|\mathcal{E}|/2}$, with $\mathcal{E}$ the set of allowable transitions $(x,y)$ with $c_{xy}\neq 0$.  
\item The set $\mathcal{J}_{\rm v}$  is the set of currents satisfying the weak symmetry Eq.~(\ref{eq:symm2}), viz.,
\begin{equation}
\mathcal{J}_{\rm v} = \left\{ J\in \mathcal{J} :  \lim_{\ell_+\rightarrow \infty}\langle T\rangle_+/\ell_+= \lim_{\ell_-\rightarrow \infty}\langle T\rangle_-/\ell_- \right\}.
\end{equation}
\item The set $\mathcal{J}_{\rm sym}$ is the set of currents that satisfy the Gallavotti-Cohen type fluctuation symmetry, 
\begin{equation}
\mathcal{J}_{\rm sym} = \left\{ J\in \mathcal{J} :  \lambda_J(a) = \lambda_J(a^\ast - a), \forall a\in \mathbb{R}\right\};
\end{equation}
\item The set $\mathcal{J}_{\rm opt}$ is the set of optimal currents,
\begin{equation}
\mathcal{J}_{\rm opt} = \left\{ J\in \mathcal{J} : \dot{s} = \overline{j}a^\ast \right\};
\end{equation}
\item The set $\mathcal{J}_S$ is the set of currents that belong to the cycle equivalence classes  $[k S_{t}]$ with $k\in \mathbb{R}$, viz., 
\begin{equation}
\mathcal{J}_{S} = \cup_{k\in \mathbb{R}}[kS_{t}].
\end{equation} 
A cycle equivalence class  is a set of currents that share the   same  cycle coefficients $c_{\gamma} = \sum_{x_j\in \gamma} c_{x_j x_{j+1}}$ for all cycles  $\gamma$  that belong to a fundamental cycle basis of the graph of admissible transitions of $X$ (see  Section 6.1 of Ref.~\cite{raghu2024effective} for details).   Thus,   the set $\mathcal{J}_S$ contains all the currents that have, up to a proportionality constant, the same cycle coefficients as the fluctuating entropy production $S_t$. 
\end{itemize}
The stars  in Fig.~\ref{fig:fig3} indicate regions in the Venn diagram  that are known to be nonempty  (as we discuss in the examples of this paper), whereas the question marks indicate regions in the Venn diagram for  which example currents have not yet been found.

\subsection{Generalised symmetries for generic currents}\label{sec:FT}
The first-passage time symmetry (\ref{eq:opt}) applies to a specific class of currents that satisfy a Gallavotti-Cohen-type fluctuation relation.  In this section we derive a generalised symmetry relation that applies to generic currents and reduces to (\ref{eq:opt}) for currents that satisfy a Gallavotti-Cohen-type fluctuation relation.   In particular, we show that the  symmetry \eqref{eq:generalisedSymmetry2} applies  to generic currents $J(t)$, i.e.,
\begin{equation}
    \mathcal{I}_-(\tau) =\hat{\mathcal{I}}^\dagger_+(\tau), \label{eq:genFlucSymIminus} 
\end{equation}
where $\hat{\mathcal{I}}^\dagger_+(\tau)$ is the rate function for the  first-passage time $T$ at the positive threshold $\ell_+$ in the time-reversal of the  dual process associated with $J$. 

The dual process is the Markov chain governed by the matrix
\begin{equation}
    \mathbf{\hat{q}} = \bm{{\phi}}_{a^\ast}^{-1} \mathbf{\tilde{q}}(a^\ast) \bm{\phi}_{a^\ast}.  \label{eq:qhat}
\end{equation}
The dual process is the Doob transform of the tilted matrix $\tilde{\mathbf{q}}(a^\ast)$~\cite{chetrite_nonequilibrium_2015} obtained by setting $a=a^\ast$  in Eq.~\eqref{eq:qtilde}.   Here, $\bm{\phi}_{a^\ast}$ is the diagonal matrix with diagonal entries given by the elements of the right eigenvector $\phi_{a^\ast}(x)$  associated with the Perron root of $\tilde{\mathbf{q}}(a^\ast)$. The dual process was identified in Ref.~\cite{neri2025martingale} as the process that determines the cumulants of $T$ at the negative threshold (although the notation   $\mathbf{q}^\ast$ was used).  Note that the   $\mathbf{\hat{q}}$ obtained from the generalised Doob transform  describes a Markov process whose trajectory weights are asymptotically equivalent to the trajectory weights of the original process $X$ conditioned on the value of the current $J(t) = t \lambda_J'(a^\ast)$ for large values of $t$~\cite{chetrite_nonequilibrium_2015}.  

The  time-reversal of the dual process is a Markov chain that is governed  by the matrix 
\begin{align}
    \mathbf{\hat{q}}^\dagger &= \bm{r}_{\rm ss}^{-1}\mathbf{\hat{q}}^{\rm T}\bm{r}_{\rm ss}\\
    &=\bm{r}_{\rm ss}^{-1}\bm{\phi}_{a^\ast} \mathbf{\tilde{q}}^{\rm T}(a^\ast) \bm{\phi}_{a^\ast}^{-1}\bm{r}_{\rm ss},\label{eq:qhatdagger}
\end{align}
where $\mathbf{r}_{\rm ss}$ is a diagonal matrix with diagonal entries given by  the steady state probability distribution $r_{\rm ss}(x)$ of the Markov chain described by $\mathbf{\hat{q}}$, and $\mathbf{r}_{\rm ss}^{-1}$ is its inverse.  

The generalised first-passage symmetry  (\ref{eq:genFlucSymIminus}) is equivalent to the following generalised fluctuation relation of the current $J$, 
\begin{equation}
    \hat{\lambda}^{\dagger}_J(a^\ast-a) = \lambda_J(a),\label{eq:genFlucSym}
 \end{equation}
where $\hat{\lambda}^{\dagger}_J(a)$ is the scaled cumulant generating function of the current $J(t)$ in the time-reversal of the  dual process.  Indeed, Eq.~\eqref{eq:genFlucSym} along with Eq.\eqref{eq:11} and Eq.~\eqref{eq:IpmLegendre} gives the generalised first passage time symmetry \eqref{eq:genFlucSymIminus}.

\begin{figure}
    \centering
    \includegraphics[width=0.8\linewidth]{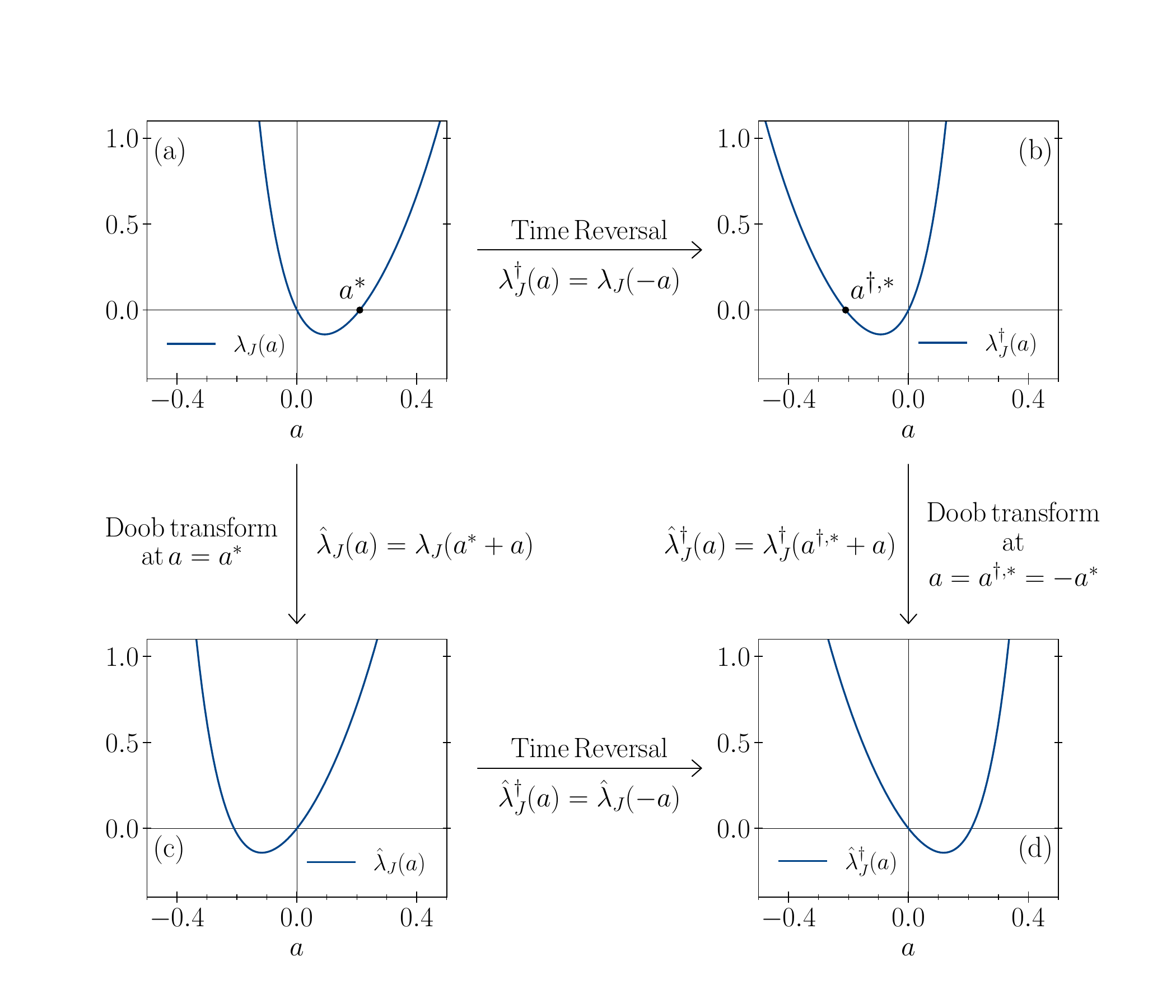}   
    \caption{ Diagram illustrating the effects of the Doob transform \eqref{eq:qhat} and  the time-reversal transformation \eqref{eq:qTimeRev} on the scaled cumulant generating function $\lambda_J$ of the current $J$.  (a) Plot of the scaled cumulant generating function $\lambda_J(a)$  in a Markov process described by the matrix $\mathbf{q}$. (b) Plot of the scaled cumulant generating function $\lambda^\dagger_J(a)$  in the  time-reversed process described by the matrix $\mathbf{q}^\dagger$ (Eq.\eqref{eq:qTimeRev}).  (c) Plot of the scaled cumulant generating function $\hat{\lambda}_J(a)$ in the  dual process described by the matrix $\mathbf{\hat{q}}$, which is the Doob-transform of $\mathbf{q}$ with tilting $a^\ast$ (Eq.\eqref{eq:qhat}).  (d) Plot of the scaled cumulant generating function $\hat{\lambda}^\dagger_J(a)$  in the conjugate process described by the matrix $\mathbf{\hat{q}^\dagger}$, which is the time-reversal of the dual process (Eq.\eqref{eq:qhatdagger}).  Note that the diagram is commutable, as one can arrive at the conjugate  process (corresponding to the plot in panel (d)) from the original process (corresponding to the plot in panel (a)) by traversing the arrows clockwise or counter-clockwise.  The values of the effective affinities $a^\ast$ and $a^{\ast,\dagger}$ in each case are also marked.}
    \label{fig:fig4}
\end{figure}

Next,  we derive the generalised fluctuation symmetry Eq.~\eqref{eq:genFlucSym}.  To do this, we consider the fluctuations of the current $J(t)$ in the Markov chain  determined by $\mathbf{\hat{q}}$. The scaled cumulant generating function $\hat{\lambda}_J(a)$  in this Markov chain is the Perron root of the  tilted matrix  of $\mathbf{\hat{q}}$, i.e., 
\begin{equation}
    \mathbf{\tilde{\hat{q}}}(a) = \mathbf{\hat{q}}\circ \exp(-a\mathbf{c}).\label{eq:qhattilde}
\end{equation}
Here, $\circ$ represents the element-wise product between two  matrices, $\mathbf{c}$ is the matrix with off-diagonal entries $c_{xy}$ and diagonal entries equal to zero, and  $\exp(-a\mathbf{c})$ is matrix whose entries are given by $\exp(-ac_{xy})$, where we have employed a slight abuse of the conventional  matrix notation for the exponential.   Using the definition of  $\mathbf{\hat{q}}$, given by Eq.~(\ref{eq:qhat}), in Eq.~(\ref{eq:qhattilde}), we obtain 
\begin{align}
    \mathbf{\tilde{\hat{q}}}(a) &=\bm{\phi}_{a^\ast}^{-1}\big[ \mathbf{\tilde{q}}(a^\ast) \circ \exp(-a\mathbf{c}) \big]\bm{\phi}_{a^\ast}\\
    &=\bm{\phi}_{a^\ast}^{-1} \mathbf{\tilde{q}}(a^\ast+a) \bm{\phi}_{a^\ast}.\label{eq:qhattildesimp}
\end{align}
Here, in  the first equality  we have used that the element-wise matrix product commutes with the diagonal transformation.  In the second equality, we have used the definition \eqref{eq:qtilde} of $\tilde{\mathbf{q}}$.  Equation~\eqref{eq:qhattildesimp} implies that  $\mathbf{\tilde{\hat{q}}}(a)$ and $\mathbf{\tilde{q}}(a^\ast+a)$ are related by a similarity transformation, and thus these two matrices have the same eigenvalues.   Hence, it holds that also their Perron roots are equal, giving
\begin{equation}
   \hat{\lambda}_J(a) =\lambda_J(a^\ast+a),\label{eq:lambdahat}
\end{equation}
as illustrated by panels (a) and (c) of Fig.~\ref{fig:fig4}. Additionally, applying Eq.~\eqref{eq:scaledTR} to  $\hat{\lambda}_J(a)$   we find that 
\begin{equation}
    \hat{\lambda}^{\dagger}_J(a) = \hat{\lambda}_J(-a),\label{eq:lamdahatdaggerlambdahat}
\end{equation}
where $\hat{\lambda}^{\dagger}_J(a)$ is the logarithmic moment generating function of $J(t)$ in the time-reversed process with rate matrix $\mathbf{\hat{q}}^\dagger$ (see panel (d) of Fig.~\ref{fig:fig4}). 
Combining Eqs.~\eqref{eq:lambdahat} and \eqref{eq:lamdahatdaggerlambdahat} we obtain  the Eq.~\eqref{eq:genFlucSym} that we were meant to show.

Note that in the above calculations there are two  operations on matrices $\mathbf{q}$:  the Doob transform  $\hat{\mathbf{q}}$ at the effective affinity $a^\ast$ [given by Eq.~(\ref{eq:qhat})], and the time-reversal $\mathbf{q}^\dagger$ [given by Eq.~(\ref{eq:qTimeRev})].   One can show that these two operations commute, i.e., $(\hat{\mathbf{q}})^\dagger = \widehat{(\mathbf{q}^\dagger)}$, as two  Markov matrices that are obtained from a diagonal transformations of a matrix have to be equal. This is illustrated in Fig.~\ref{fig:fig4} in terms of the scaled cumulant generating function of the current $J(t)$ in each of the processes described by the matrices $\mathbf{q}$, $\hat{\mathbf{q}}$,  $\mathbf{q}^\dagger$, and $\mathbf{\hat{q}^\dagger}$.

\section{Examples}\label{sec:examples}
We illustrate the bounds on dissipation and the symmetry properties of currents on two toy examples of nonequilibrium processes. 

The right-hand sides of the inequalities (\ref{eq:unc}) and (\ref{eq:totalbound}) determine the amount of the time-irreversibility, and thus also dissipation, captured by the random variables $T$ and $D$.    The right-hand side of (\ref{eq:unc}), 
\begin{equation}
\hat{s}_{\rm FPR} = \frac{\ell_+}{\ell_-} \frac{|\ln p_-|}{\langle T\rangle}, \label{eq:sfpr}
\end{equation}
has been referred to as the first-passage ratio~\cite{neri2022estimating,neri2023extreme}.  The first-passage ratio
 determines how much of the dissipation $\dot{s}$ in the process $X$ is captured by the random variable $D = {\rm sign}(J(T))$. The right-hand side of (\ref{eq:totalbound}) considers the amount of dissipation captured by both $D$ and the first-passage time $T$.  We call   the right-hand  side of  (\ref{eq:totalbound}) the improved first-passage ratio, 
\begin{equation}
\hat{s}_{\rm iFPR} = \frac{\ell_+}{\langle T\rangle} \left(\frac{|\ln p_- |}{\ell_-} + \mathcal{I}_-\left(1/\overline{j}\right)\right). \label{eq:sifpr}
\end{equation}
The difference  $\hat{s}_{\rm iFPR}-\hat{s}_{\rm FPR}$  determines the additional dissipation  captured by the first-passage time $T$ that is not captured  by $D$. 

In the following we  determine $\hat{s}_{\rm FPR} $ and $\hat{s}_{\rm iFPR}$ for two toy models, both of which are  Markov chains in continuous time.  In this way, we will quantify the amount of the dissipation contained in the time-irreversibility of the random variables $D$ and $T$.   Furthermore, we will find examples of   currents that satisfy the symmetry condition (\ref{eq:weaksymm}) and discuss their place in the Venn diagram of Fig.~\ref{fig:fig3}. 

\subsection{Biased random walker on a two-dimensional lattice}\label{sec:2DRandomd}
We  consider  a random walker that moves on a two-dimensional lattice with periodic boundary conditions (equivalent to a lattice on a torus)~\cite{neri2022estimating}.    The process $X = (X_1,X_2)$  denotes the position of the random walker, with $X_1$ and $X_2$ the two spatial coordinates of $X$.   The coordinate processes are  two independent jump process  described by 
\begin{equation}
\mathrm{d} X_i(t)=\mathrm{d} N_i^{+}(t)-\mathrm{d} N_i^{-}(t), \quad i \in\{1,2\},
\end{equation}
where $N_i^+$ and $N^-_i$ are  Poisson counting processes with rates $k_i^+$ and $k^-_i$,  respectively.  
We parametrise the rates of the jump processes as 
\begin{eqnarray}
    k_1^+ = \frac{\exp(\nu/2)}{4\cosh(\nu/2)}, &\quad & k_1^- =\frac{\exp(-\nu/2)}{4\cosh(\nu/2)},\\
    k_2^+ = \frac{\exp(\rho\nu/2)}{4\cosh(\rho\nu/2)}, &\quad& k_2^- =\frac{\exp(-\rho\nu/2)}{4\cosh(\rho\nu/2)},
\end{eqnarray}    
where $\nu$ and $\rho$ are two parameters controlling the bias of  the motion in the  two directions.  The parametrisation sets  $k^+_1 + k_1^- + k_2^+ + k_2^-=1$, corresponding to a choice of time units such that the average time for the particle to move equals one.  Moreover, this parametrisation sets $k^+_1+k^-_1 = k^+_2+k^-_2$ so that the mean jump time is the same in both directions. 

Up to an irrelevant scaling constant, an arbitrary current in this model may be expressed  as
\begin{equation}
    J(t) = (1-\Delta) J_1(t)+ (1+\Delta) J_2(t) \label{eq:JDef}
\end{equation}
in terms of a single parameter, $\Delta\in \mathbb{R}$,
where $J_i(t) = N_i^+(t)-N_i^-(t)$, with $i\in{1,2}$, are the currents in the two directions.

The average rate of entropy production for the random walker is
\begin{equation}
    \dot{s} = \nu(k_1^+ - k_1^-) + \nu\rho(k_2^+-k_2^-)
\end{equation}
and the fluctuating entropy production equals 
\begin{equation}
    S(t) = \nu J_1(t) + \rho\nu J_2(t).
\end{equation}
The fluctuating entropy production $S$ is proportional to $J$  if $\Delta = (1-\rho)/(1+\rho)$.

\begin{figure}
    \centering
     \includegraphics[width=0.5\textwidth]{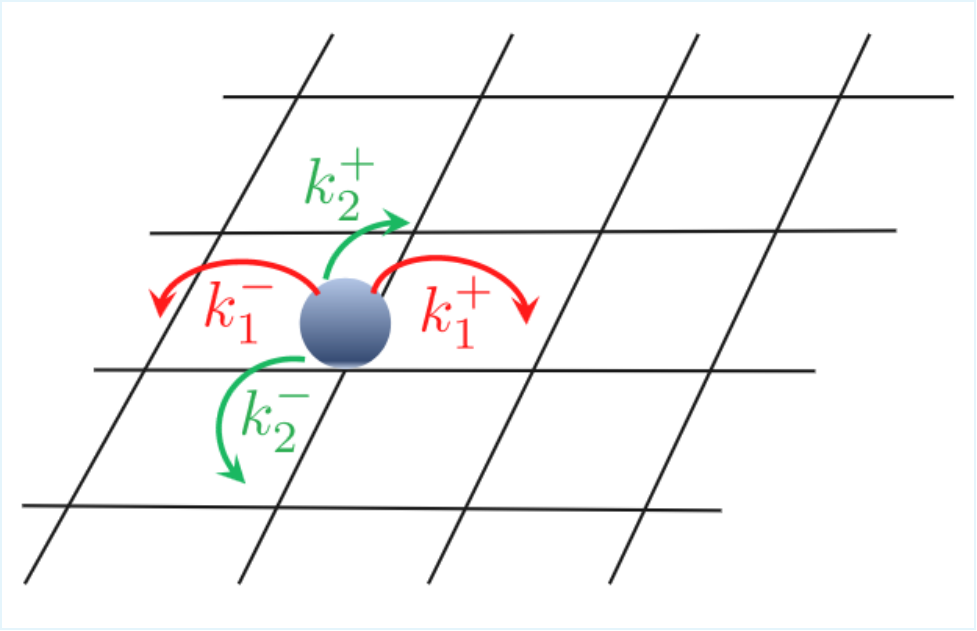}
    \includegraphics[width=1.05\linewidth]{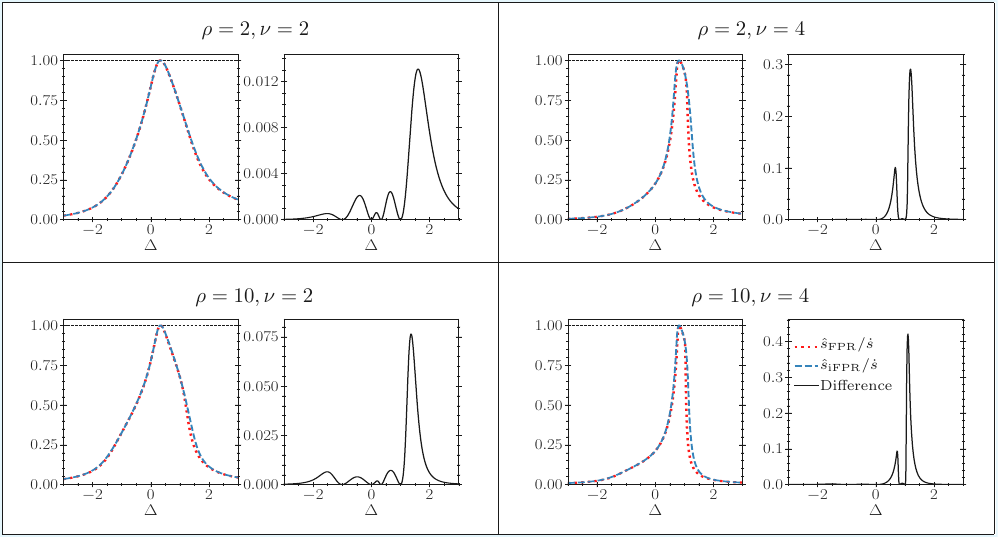}
    \caption{ Top Panel: Graphical illustration of the random walker model on a two-dimensional lattice (Figure is taken from Ref.~\cite{neri2022estimating}). Bottom Panel:  Plots of the ratios $\hat{s}_{\rm FPR}/\dot{s}$ (dotted, red line),  $\hat{s}_{\rm iFPR}/\dot{s}$  (dashed, blue line), and $(\hat{s}_{\rm iFPR}-\hat{s}_{\rm FPR})/\dot{s}$ (black, solid line) as a function of $\Delta$ for the two-dimensional random walker model as defined in Sec.~\ref{sec:2DRandomd}.  The model parameters  $\rho$ and $\nu$ are inidicated  in the figure.   The horizontal dashed line indicates the constant function  $1$.  See~\ref{app:D1} for a description of the numerics used  to generate the plots.}
    \label{fig:fig5}
   \end{figure}

Figure~\ref{fig:fig5} shows the ratios $\hat{s}_{\rm FPR}/\dot{s}$ and $\hat{s}_{\rm iFPR}/\dot{s}$ as a function of the parameter $\Delta$ that specifies the current $J$ for four  processes $X$ corresponding with distinct values of $\rho$ and $\nu$ (left panels of the four subfigures).    These plots reveal that the random variable $D$  captures for a broad range of parameters $\Delta$  a large fraction  of the total dissipation $\dot{s}$, up to $100\%$ of the total dissipation when $\Delta = (1-\rho)/(1+\rho)$.  Instead, the random variable $T$ in general  captures a small fraction of the dissipation $\dot{s}$.    This is more clearly visible in the right panels of the four subfigures, where $(\hat{s}_{\rm FPR}-\hat{s}_{\rm iFPR})/\dot{s}$  is plotted as a function of $\Delta$.   We observe that for most examples of currents the fraction of dissipation captured by $T$ is negligible.   Nevertheless,  as shown in the panels with $(\rho,\nu) = (2,4)$ and $(\rho,\nu) = (10,4)$,  there do exist currents for which $T$ captures up to $40\%$ of the total dissipation.   These are currents for which the statistics of first-passage times $T$ is highly asymmetrical between both thresholds.

From the right-panels of the Figs.~\ref{fig:fig5}, we can identify the currents for which $\hat{s}_{\rm FPR}-\hat{s}_{\rm iFPR} = 0$, and thus the symmetry condition (\ref{eq:weaksymm}) is satisfied.  The relevant currents are those with: (i) $\Delta= (1-\rho)/(1+\rho)$, which is a current proportional to $S(t)$.  This current  belongs to the set $\mathcal{J}_S$ and is an example of an optimal current. (ii)   Currents with $\Delta\in \left\{-1,0,1\right\}$ that satisfy the Gallavotti-Cohen-type fluctuation symmetry but are not optimal.  These currents belong to the set  $\mathcal{J}_{\rm sym}\setminus \mathcal{J}_{S}$.    Hence,  in this model all  currents that satisfy the weak symmetry condition 
(\ref{eq:weaksymm}) also satisfy the stronger symmetry condition (\ref{eq:opt}).

\subsection{Markov jump process with four states}\label{sec:fourstate}
\begin{figure}
    \centering
    \includegraphics[width=0.87\linewidth]{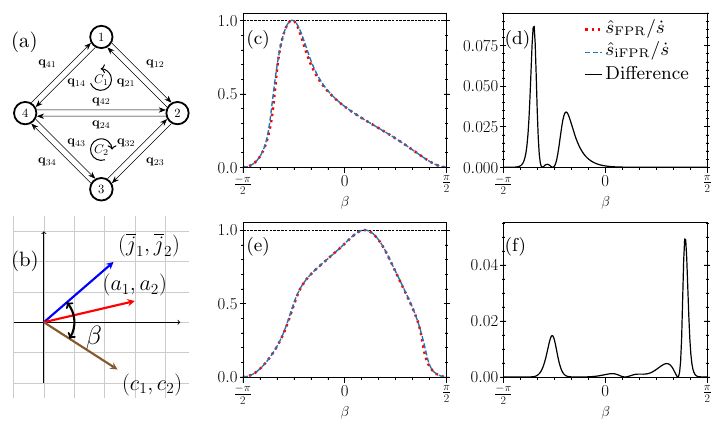}
    \caption{
    (a) Graphical illustration of four state model being studied: the graph of admissible transitions for the four state model studied in Sec~\ref{sec:fourstate} is depicted, with each transition labelled with the corresponding jump rate. Also indicated are the cycles $C_1$ and $C_2$ of the chosen cycle basis, along with their respective directions. (b) Graphical illustration of parameters in four state model: the cycle currents $(\overline{j}_1,\overline{j}_2)$, the cycle coefficients $(c_1,c_2)$ and the cycle affinities $(a_1,a_2)$ are plotted as vectors in $\mathbb{R}^2$ for an example system and current $J(t)$. The angle $\beta$ which determines the estimators  $\hat{s}_{\rm FPR}/\dot{s}$ and $\hat{s}_{\rm iFPR}/\dot{s}$ for the current $J$ is indicated.
    (c),(d) Plots of the ratios $\hat{s}_{\rm FPR}/\dot{s}$ (dotted, red line),  $\hat{s}_{\rm iFPR}/\dot{s}$  (dashed, blue line), and $(\hat{s}_{\rm iFPR}-\hat{s}_{\rm FPR})/\dot{s}$ (black, solid line) as a function of $\beta$ for the system depicted in panel (a) and randomly generated jump rates as described in~\ref{app:D2}. (e),(f) Plots of the ratios $\hat{s}_{\rm FPR}/\dot{s}$ (dotted, red line),  $\hat{s}_{\rm iFPR}/\dot{s}$  (dashed, blue line), and $(\hat{s}_{\rm iFPR}-\hat{s}_{\rm FPR})/\dot{s}$ (black, solid line) as a function of $\beta$ as described before for a different randomly generated jump rates as specified in~\ref{app:D2}}
    \label{fig:fig6}
\end{figure}

We illustrate the application of the bounds \eqref{eq:unc} and \eqref{eq:totalbound} on a simple  Markov jump process for which its  graph of admissible transitions is illustrated in   Panel (a)  of Fig.~\ref{fig:fig6}.  This Markov jump process 
 has four states, i.e., $X(t)\in \mathcal{X}=\left\{1,2,3,4\right\}$, and the graph of admissible transitions has  two independent fundamental cycles $C_1 = (1,4,2,1)$ and $C_2 = (2,4,3,2)$  [as indicated in Fig.\ref{fig:fig6}(a)]. Note that this process is arguably the simplest example of a Markov chain  that has non-optimal currents, as  for unicyclic systems  all currents are optimal \cite{raghu2024effective}. We randomly generate the jump rates $\mathbf{q}_{xy}$ as described in~\ref{app:D2}. 

In this example, for a given set of rates $\mathbf{q}$, the  first-passage ratios  $\hat{s}_{\rm FPR}$ [given by~\eqref{eq:sfpr}] and $\hat{s}_{\rm iFPR}$~[given by \eqref{eq:sifpr}] 
for the different stochastic currents $J(t)$  are determined by a single parameter $\beta$ (see Sec.7 of \cite{raghu2024effective}).   The parameter $\beta$ is the angle between the \textit{cycle coefficients} $(c_1,c_2)$ associated with the current $J(t)$ and the \textit{cycle currents} $(\overline{j}_1,\overline{j}_2)$ of the Markov process $X$, when these are plotted as vectors in $\mathbb{R}^2$.  This is illustrated in Panel (b) of Fig.~\ref{fig:fig6}. Here, the cycle coefficients $c_1$ and $c_2$ are defined as  the sum of the coefficients $c_{xy}$ along the cycles $C_1$ and $C_2$, respectively, i.e., 
\begin{equation}
    c_i = \sum_{x_j \in C_i} c_{x_jx_{j+1}} \quad {\rm for}\,\, i=1,2, 
\end{equation}
where $c_{x,y}$ are the coefficients used to define the current $J(t)$ in Eq.\eqref{eq:J}. 
The cycle currents $(\overline{j}_1,\overline{j}_2)$ are the average currents associated with the cycles $C_1$ and $C_2$ such that the average  of any current $J$ takes the form
\begin{equation}
    \overline{j} = c_1\overline{j}_1 +c_2\overline{j}_2.\label{eq:jbarcycles}
\end{equation}
Note that the values of 
$\overline{j}_1$ and $\overline{j}_2$ are uniquely determined by the matrix $\mathbf{q}_{xy}$ and the  choice of the cycle basis $(C_1,C_2)$.   As shown in  Sec.~6.1 of Ref.~\cite{raghu2024effective}, the rates  $\overline{j}_1$ and $\overline{j}_2$ can be expressed in terms of   the average edge currents 
\begin{equation}
    \overline{j}_{xy} = \lim_{t\to\infty}   \frac{\langle N^{xy}(t)\rangle  - \langle N^{yx}(t)\rangle }{t}  
\end{equation}
between  pairs of states $x,y\in\mathcal{X}$. We introduce the equation
\begin{equation}
    \overline{j}_{xy} = \eta_1 j_1 + \eta_2 j_2,\label{eq:etaj}
\end{equation}
where, for $i \in \{1,2\}$, $\eta_i=0$ if the edge $xy$ is not in the cycle $C_i$ and $\eta_i=+1$ ($-1$) if the edge $xy$ is in the cycle $C_i$ and oriented along (against) the direction of the cycle as a requirement on the currents $j_1$ and $j_2$.  Solving Eq. \eqref{eq:etaj}  together with    Kirchoff's laws  towards $j_1$ and $j_2$, we obtain   
\begin{equation}
    \overline{j}_1 = \overline{j}_{14} = \overline{j}_{21} \quad {\rm and }\quad \overline{j}_2=\overline{j}_{23} = \overline{j}_{34},
\end{equation}
which satisfy Eqs. \eqref{eq:jbarcycles} and \eqref{eq:etaj}.
Panel (b) of Fig.\ref{fig:fig6} also indicates the \textit{cycle affinities}~\cite{bauer2014affinity}, 
\begin{equation}
    a_i = \sum_{x_j \in C_i} \ln\frac{\mathbf{q}_{x_jx_{j+1}}}{\mathbf{q}_{x_{j+1}x_{j}}} \quad {\rm for}\,\, i=1,2, 
\end{equation}
Currents $J$ for which $(c_1,c_2)$ is parallel to $(a_1,a_2)$ are   optimal. 

We clarify why the functional dependency of the ratios  $\hat{s}_{\rm FPR}$ and $\hat{s}_{\rm iFPR}$ on the coefficients $c_{xy}$ can be reduced to a functional dependency on  a single parameter $\beta$.   This follows from the fact that $\hat{s}_{\rm FPR}$ and $\hat{s}_{\rm iFPR}$  are determined by  the scaled cumulant generating function $\lambda_J$, and it was shown in Ref.~\cite{raghu2024effective} that $\lambda_J$ is a function of the cycle coefficients $(c_1,c_2)$.   Further, both bounds are invariant under a rescaling  the current  $J$ by a constant  (i.e., by multiplying all coefficients $c_{xy}$ by a fixed real number).  Hence, all possible values of the  ratios  $\hat{s}_{\rm FPR}$ and $\hat{s}_{\rm iFPR}$  can be plotted as a function of $\beta\in[-\pi/2,\pi/2)$.

Panels (c) and (e) of Fig.~\ref{fig:fig6}  show the ratios $\hat{s}_{\rm FPR}/\dot{s}$ and $\hat{s}_{\rm iFPR}/\dot{s}$ as a function of $\beta$ for two randomly generated rate matrices $\mathbf{q}$ with a graph of admissible transitions as indicated in  Panel (a) of Fig.~\ref{fig:fig6}.     The results are in agreement with  those of the two-dimensional random walker consider in the previous section, in the sense that the first-passage time $T$ contains a small amount of information on the arrow of time.   

Panels (d) and (f) of Fig.~\ref{fig:fig6} plot the corresponding differences between the two ratios, $(\hat{s}_{\rm iFPR}-\hat{s}_{\rm FPR})/\dot{s}$ as a function of $\beta$.  We observe that this ratio equals zero when $(c_1,c_2)$ is parallel to $(a_1,a_2)$, corresponding with the optimal case.    Furthermore, we  observe that there exist other currents for which the difference between the two ratios vanishes, and thus for which the  symmetry relation (\ref{eq:weaksymm}) holds.   Interestingly, these currents do not satisfy the Gallavotti-Cohen symmetry and hence they belong to the set $\mathcal{J}_{\rm v}\setminus \mathcal{J}_{\rm sym}$ in the Venn diagram of Fig.~\ref{fig:fig3} (see \ref{app:F} for more details).

\section{Discussion}\label{sec:disc} 
We elaborate here on a few remaining points that we think are worth discussing.In Sec.\ref{sec:Disc1}, we summarise the properties of the  paper's setup in Sec.~\ref{sec:setup} that we   used to derive the main results, and we   also discuss how the results can be extended beyond the  setup of the paper; in Sec.\ref{sec:Disc2}, we examine the differences between first-passage problems in discrete and continuous time; and lastly, in Sec.~\ref{sec:Disc3}, we explore potential applications of the paper’s main results in physical chemistry.

\subsection{Extensions}\label{sec:Disc1}
The  inequalities (\ref{eq:unc})  and  (\ref{eq:totalbound}) rely on the following properties of the setup  in Sec.~\ref{sec:setup}:  we require that (i) the thresholds $\ell_-$ and $\ell_+$ are infinitely large; (ii) the process $X_t$ is ergodic; (iii) the process $X_t$   has even parity  under time-reversal, i.e., the time reversed Markov chain is described by Eq.~(\ref{eq:qTimeRev}); (iv) the time-additive observable is a fluctuating current, i.e., $c_{xy}=-c_{yx}$, and its average rate $\overline{j}$ is positive, $\overline{j}>0$; and (v) the set $|\mathcal{X}|$ is finite.

The first four properties (i)-(iv) of the setup are essential for  deriving Eqs.~(\ref{eq:unc})  and  (\ref{eq:totalbound}).  The property (v) is less essential, as  the inequalities Eqs.~(\ref{eq:unc})  and  (\ref{eq:totalbound})   apply in general to Markov jump processes defined on sets $\mathcal{X}$ of infinite cardinality or for overdamped Langevin processes.   Nevertheless, some of the results — such as Eqs.~(\ref{eq:pDagger}) — were derived under the assumption of finite cardinality. Therefore, caution should be exercised when extending these results to cases with infinite cardinality.  We will discuss this scenario in more detail in a future work.

For  setups that do not satisfy the properties (iii) and (iv),  the main results of this paper do not hold.  However, the following  more general inequalities   apply (excluding cases with $p^\dagger_+=0$)
 \begin{equation}
\dot{s}  \geq \frac{ | \ln  p^\dagger_+|}{\langle T\rangle} (1+o_{\ell_{\rm min}}(1)), \label{eq:timeRevTradeoffv2}
 \end{equation}
 and 
\begin{equation}
\dot{s}\geq  \frac{\ell_+}{\langle T\rangle} \left(\frac{|\ln p^\dagger_+|}{\ell_-} +   \mathcal{I}^\dagger_+(1/\overline{j}) \right)(1+o_{\ell_{\rm min}}(1)). \label{eq:boundOv2}
\end{equation} 
For example, Eqs.~(\ref{eq:timeRevTradeoffv2})  and (\ref{eq:boundOv2}) apply to time-additive observables that are not fluctuating currents, such as the total number of jumps towards a state $x$ minus the number of jumps towards another state $y$~\cite{neri2025martingale}, or to fluctuating currents in systems with  magnetic fields~\cite{chun2019effect}. 
The assumptions (iii) and (iv) are crucial for the derivation of Eqs.(\ref{eq:unc}) and (\ref{eq:totalbound}) because, without them, the symmetries expressed by Eqs.(\ref{eq:pDagger}) and (\ref{eq:ITR}) do not hold. As a result, we cannot relate the time-reversed quantities $p^\dagger_+$ and $\mathcal{I}^\dagger_+$ to their counterparts in the forward dynamics.      Instead,   the  main limitations on the  validity   of  the  general inequalities (\ref{eq:timeRevTradeoffv2}) and (\ref{eq:boundOv2})   stem from the assumptions required to derive Wald’s equality $\langle S(T)\rangle  = \dot{s}\langle T\rangle (1+o_{\ell_{\rm min}}(1))$, namely, large thresholds, the ergodicity of the process $X$, and the fact that $T$ is a stopping time (a random time that obeys causality).

The inequalities Eqs.~(\ref{eq:timeRevTradeoffv2})  and (\ref{eq:boundOv2}), although less useful for  the inference of kinetic properties of molecular systems,  clarify some of the phenomenology   on trade-offs between dissipation and accuracy as observed in the literature.  These inequalities imply that  the   trade-off between dissipation  and accuracy should be understood  as a trade-off with the accuracy in the time-reversed process. The fact that for some systems this trade-off also involves the accuracy of the original process arises from the coincidence that in these systems the statistics of the forward and reverse processes are identical.    
An illustrative example is the dynamics of a charged particle moving in a magnetic field and driven out of equilibrium by an external torque. As shown in Ref.~\cite{chun2019effect},  increasing the magnetic field leads to  a decrease in the rate of dissipation and an increase in the accuracy, and thus remarkably there  is no trade-off between dissipation and accuracy. Nevertheless, according to Eq.~(\ref{eq:boundOv2}), in the time-reversed process, where the sign of the magnetic field is inverted, accuracy should correspondingly increase. Therefore, the observed absence of a trade-off between dissipation and accuracy in  systems with magnetic fields stems from the fact that the time-reversed dynamics differs from the forward dynamics, and thus accuracy should be measured in the system with the magnetic field inverted to observe the trade-off.

\subsection{Discrete versus continuous time}\label{sec:Disc2}
The derivations in this paper also clarify  distinctions  and similarities between the first-passage time statistics of currents in discrete-time and continuous-time Markov chains.

The inequalities   (\ref{eq:unc})  and (\ref{eq:totalbound})  apply to both discrete-time and continuous-time Markov chains. This was not evident from previous derivations~\cite{neri2022universal, raghu2024effective}, which relied on the parabolic bound on the scaled cumulant generating function of $J(t)$\cite{pietzonka, gingrich2016dissipation},
\begin{equation}
    \lambda_{J}(a) \geq a \bar{j}\left(-1+\frac{a \bar{j} }{ \dot{s}}\right),\label{eq:lambdaPB}
\end{equation}
a result specific to continuous-time Markov chains~\cite{proesmans2017discrete}.        Interestingly,  this   implies that the effective affinity, $a^\ast$, which serves as a generalization of chemical affinity to systems with multiple coupled currents ~\cite{raghu2024effective}, also extends to  discrete-time Markov chains. 
 
To guarantee the existence of $a^\ast$ in the discrete time case, we need to demonstrate that   $\lambda_J(a)$ has a nonzero root. This can be seen by mapping the discrete time Markov chain governed by the transition matrix $\mathbf{q}$ to a continuous time Markov process governed by a transition matrix $\mathbf{q}_c= \mathbf{q} - \mathds{1}$, where $\mathds{1}$ is the identity matrix, as described in \cite{chiuchiu_mapping_2018}. The continuous time process governed by $\mathbf{q_c}$ is now one for which the effective affinity $a^\ast$ exists (as implied by the parabolic bound \eqref{eq:lambdaPB}). We then note that the root $a^\ast$ of $\lambda_J(a)$ in the continuous time case must also be a root of $\lambda_J(a)$ in the corresponding discrete time Markov chain (as can be seen from Eqs. (12) and (14) of \cite{chiuchiu_mapping_2018}).

Note that the thermodynamic uncertainty relation (\ref{eq:TUR})  for first-passage times does not apply to discrete-time Markov chains.   Nevertheless,  our analysis shows that the following  discrete-time thermodynamic uncertainty relation holds for first-passage times in discrete time,
\begin{equation}
\exp(\dot{s})-1 \geq 2 \frac{\langle T\rangle_+}{  \langle T^2\rangle_+-\langle T\rangle^2_+},  \label{eq:proes}
\end{equation}
where, as before, we have assumed that the time unit between two consecutive time-steps  equals one.   The inequality (\ref{eq:proes}) follows from the inequality (2) of Ref.~\cite{proesmans2017discrete} and the identities 
\begin{equation}
\overline{j} =  \lim_{\ell_+\rightarrow \infty}\frac{\ell_+}{\langle T\rangle_+}
\end{equation}
and 
\begin{equation}
\lim_{t\rightarrow \infty}\frac{\langle J^2(t)\rangle - \langle J(t)\rangle^2 }{t} =  \overline{j}^3\lim_{\ell_+\rightarrow \infty}\frac{\langle T^2\rangle_+  - \langle T\rangle^2_+}{\ell_+}
\end{equation}
that can be derived from Eqs.~(\ref{eq:muPmuM}) by following derivations similar to those in continuous time Markov chains  \cite{neri2025martingale}.  

We summarise in Table~\ref{tab:tab1} the results we discussed in this paper and their validity in continuous and discrete time cases.

\begin{table}[h!]
    \centering
    \begin{TAB}(r,0.5cm,1cm)[5pt]{|c|c|c|c|}{|c|c|c|c|c|c|c|c|c|c|}
        Equation & Number & Continuous-Time & Discrete-Time \\
        $\dot{s} \geq \frac{\ell_{+}}{\ell_{-}} \frac{\left|\ln p_{-}\right|}{\langle T\rangle}\left(1+o_{\ell_{\min }}(1)\right)$ & \eqref{eq:unc} & \checkmark & \checkmark \\
        $\dot{s} \geq \bar{j} a^\ast$ & \eqref{eq:sjas} & \checkmark & \checkmark \\
        $\dot{s} \geq \frac{\ell_{+}}{\langle T\rangle}\left(\frac{\left|\ln p_{-}\right|}{\ell_{-}}+\mathcal{I}_{-}(1 / \bar{j})\right)\left(1+o_{\ell_{\min }}(1)\right)$ & \eqref{eq:totalbound} & \checkmark & \checkmark \\
        $\dot{s} \geq I_J(-\bar{j})=\tilde{a} \bar{j}-\lambda_J(\tilde{a})$ & \eqref{eq:newboundIJform} & \checkmark & \checkmark \\
        $\dot{s} \geq \frac{\left|\ln p_{+}^{\dagger}\right|}{\langle T\rangle}\left(1+o_{\ell_{\min }}(1)\right)$ & \eqref{eq:asymptotic} & \checkmark & \checkmark \\
        $\left|\ln p_{+}^{\dagger}\right|=\frac{\ell_{+}}{\ell_{-}}\left|\ln p_{-}\right|\left(1+o_{\ell_{\min }}(1)\right)$ & \eqref{eq:pDagger} & \checkmark & \checkmark \\
        $\lambda_{J}(a) \geq a \bar{j}(-1+a \bar{j} / \dot{s})$ & \eqref{eq:lambdaPB} & \checkmark & $\times$ \\
        $\dot{s} \geq \frac{2}{\langle T\rangle_{+}} \frac{\langle T\rangle_{+}^2}{\left\langle T^2\right\rangle_{+}-\langle T\rangle_{+}^2}\left(1+o_{\ell_{+}}(1)\right)$ & \eqref{eq:TUR} & \checkmark & $\times$ \\
        $\exp (\dot{s})-1 \geq \frac{2}{\langle T\rangle_{+}} \frac{\langle T\rangle_{+}^2}{\left\langle T^2\right\rangle_{+}-\langle T\rangle_{+}^2}\left(1+o_{\ell_{+}}(1)\right)$ & \eqref{eq:proes} & \checkmark & \checkmark \\
    \end{TAB}
    \caption{We summarize the validity of results discussed in this paper in discrete and continuous time. Note that the majority of the results that we derive in this paper hold for both cases, with the exception of the inequalities \eqref{eq:lambdaPB}  and (\ref{eq:TUR}) that do not apply in discrete time Markov chains.}
    \label{tab:tab1}
\end{table}

\subsection{Thermodynamic inference from first-passage time statistics  }\label{sec:Disc3}
The results of this paper can be applied to infer the kinetic and thermodynamic properties of macromolecular systems from observations of a fluctuating current $J$, as well as, to develop coarse-grained descriptions of their dynamics.    A good example are molecular motors, where $J$ corresponds to the position, and $X$ represents the different conformational and chemical states of the motor. In this context, experimental results on the fluctuations of $D ={\rm sign}(J(T))$ and  $T$ are available~\cite{nishiyama2002chemomechanical,asbury2003kinesin,rief2000myosin}.   

The experiments on kinesin in  Ref.~\cite{nishiyama2002chemomechanical}  show that   for molecular motors $\langle T\rangle_+\approx \langle T\rangle_-$ (see  Panel b  of  Figure 5 in Ref.~\cite{nishiyama2002chemomechanical}).    Based on our  results we can conclude that  are two possible interpretations for this symmetry.     Either  (i) the  positional current is  optimal  (it belongs to $\mathcal{J}_{\rm opt}$ in Fig.~\ref{fig:fig3}), in which case $a^\ast \overline{j} \approx \dot{s}$; or  (ii) the positional current  belongs to the set  $\mathcal{J}_{\rm v}\setminus\mathcal{J}_{\rm opt}$, and thus the positional current satisfies 
$\langle T\rangle_+\approx \langle T\rangle_-$ even though it is not optimal.    Note that the symmetry in the first-passage time of $J$ is relevant for developing thermodynamically consistent coarse-grained of molecular motors~\cite{wang2007detailed}. 

The inequalities (\ref{eq:unc}) and (\ref{eq:totalbound}) can be used to bound the  efficiency of the chemomechanical coupling in motor proteins based on measurements of the fraction of backsteps (through $p_-$) and the fluctuations in the dwell times $T$ (through $\mathcal{I}_-(1/\overline{j})$)~\cite{neri2022estimating}.   
In this regard, the examples in Sec.~\ref{sec:examples} show that in general the improvement gained from  the first-passage times $T$ is  small.   It remains to be understood, also in the perspective of thermodynamically consistent coarse-graining~\cite{teza2020exact,igoshin2025coarse}, what observable contains the part of $\langle S(T)\rangle$ that is not already contained in $D={\rm sign}(J(T))$.

Lastly, let us compare the inequalities (\ref{eq:unc}) and (\ref{eq:totalbound}) with other approaches in the literature that bound  the rate of dissipation based on partial observations in a Markov process~\cite{harunari_what_2022,partial,blom2024milestoning,ertel2024estimator,fritz2025entropy}.  
The inequality (\ref{eq:totalbound}) has a formal similarity with the "waiting-time" bounds of Refs.~\cite{harunari_what_2022,partial}, which also consist of two terms, one with and one without  temporal information   (for example, see  Eq.(6) of~ Ref.~\cite{harunari_what_2022}).    Both approaches have advantages and disadvantages.   Approaches as in
  Refs.~\cite{harunari_what_2022,partial} require that the observer can measure transitions between individual  states of the Markov process $X$.   It is unclear how strong this approximation is for macromolecular systems, as  simple proteins, such as myoglobin, have a  very large number of  
quasi-degenerate microscopic states~\cite{frauenfelder1991energy}.  Hence, a very good microscope would be required to measure the individual transitions.  Instead, the approach in the present paper, based on the inequalities  (\ref{eq:unc}) and (\ref{eq:totalbound}), does not assume that the observer can measure  transitions between individual states of  $X$.   Instead, an observer just needs to be able to measure a fluctuating current, which is inline with experimental setups that measure the positional current of a molecular motor~\cite{nishiyama2002chemomechanical}.   However, a drawback of the present approach is that we require the limit of large thresholds.   This drawback may be less significant than it initially appears: case studies indicate that the asymptotic limit of large thresholds is rapidly reached when the system is far from thermal equilibrium, as in this regime  $p_-$  describes rare events even at low thresholds. 
Overall, the different approaches are based on distinct approximate representations of  experimental setups. Therefore, it is likely prudent to employ a number of different methods when inferring thermodynamic and kinetic properties from the fluctuations of a mesoscopic system.  

\appendix

\section{Coarse-graining of $\langle S(T)\rangle$: general approach from a probability theoretic perspective}\label{app:conv}
The coarse-graining of $\langle S(T)\rangle$ that we use in Secs.~\ref{sec:firstbound}  and \ref{sec:PT}   
applies generically to the  average of the entropy production $S$   evaluated at a stopping time $T$ of a stochastic process $X$.   As we discuss in this appendix, this  follows from the fact that $\langle S(T)\rangle$  can  be expressed a Kullback-Leibler divergence, and  coarse-graining applies generically        to Kullback-Leibler divergences.  We use here a general probability theoretic approach, which relies on concepts developed in the theory of  sequential hypothesis testing~\cite{tartakovsky2014sequential}.    For terminology we refer to text books in probability theory~\cite{williams1991probability,liptser2001statistics}.     

\subsection{Defining $\langle S(T)\rangle$ in a probability theoretical fashion }

Consider a filtered probability space $(\Omega,\mathcal{F},\left\{\mathcal{F}_t\right\}_{t\geq 0},\mathbb{P})$, where $\Omega$ is the set of outcomes, $\mathcal{F}$ is a $\sigma$-algebra of $
\Omega$,  $\left\{\mathcal{F}_t\right\}_{t\geq 0}$ is an increasing sequence of sub-$\sigma$-algebras ($\mathcal{F}_{s}\subset \mathcal{F}_t$ for all $t>s$), and $\mathbb{P}$ is a probability measure defined on $\mathcal{F}$.   Furthermore, we consider a second probability measure $\mathbb{P}^\dagger$ so that  $\mathbb{P}$ and $\mathbb{P}^\dagger$ are locally, mutually, absolutely continuous.   This latter means that $\mathbb{P}(\Phi) = 0 \Leftrightarrow \mathbb{P}^\dagger(\Phi)=0$ for all $\Phi\in \mathcal{F}_t$ and all $t\geq 0$.  

For the construction of the filtered probability space $(\Omega,\mathcal{F},\left\{\mathcal{F}_t\right\}_{t\geq 0},\mathbb{P})$  of a Markov chain we refer to Refs.~\cite{norris1998markov, liggett2010continuous}.   Briefly, for Markov chains the elements of $\Omega$ are the trajectories $x^\infty_0$.   The sub-$\sigma$-algebras  $\mathcal{F}_t$ are those generated by the random variables $X(t')$ with $t'\in [0,t]$.   The probability measure $\mathbb{P}$ on $\mathcal{F}$ is constructed using  the statistics provided by the $\mathbf{q}$-matrix  and the probability mass function  $p_{{\rm init}}$ of the initial state.

We define the  stochastic entropy production as 
\begin{equation}
S(t) = \ln \frac{\left.d\mathbb{P}\right|_{\mathcal{F}_t}}{\left.d\mathbb{P}^\dagger\right|_{\mathcal{F}_t}},
\end{equation}
where 
\begin{equation}
\frac{\left.d\mathbb{P}^\dagger\right|_{\mathcal{F}_t}}{\left.d\mathbb{P}\right|_{\mathcal{F}_t}} = e^{-S(t)}
\end{equation}
is the Radon-Nikodym derivative of the measure $\left.\mathbb{P}^\dagger\right|_{\mathcal{F}_t}$ with  respect to $\left.\mathbb{P}\right|_{\mathcal{F}_t}$.    Here, the measures $\left.\mathbb{P}\right|_{\mathcal{F}_t}$ and  $\left.\mathbb{P}^\dagger\right|_{\mathcal{F}_t}$ are the restrictions of the measures  $\mathbb{P}$ and $\mathbb{P}^\dagger$ with respect to the filtration $\mathcal{F}_t$.   This means that  $\left.\mathbb{P}\right|_{\mathcal{F}_t}(\Phi) = \mathbb{P}(\Phi)$ for all $\Phi\in \mathcal{F}_t$ and $\left.\mathbb{P}\right|_{\mathcal{F}_t}(\Phi) = 0$ for all $\Phi\in \mathcal{F} \setminus \mathcal{F}_t$.   

Let $T$ be a stopping time, i.e., a random time $T$  for which $\left\{T\leq t\right\}\in \mathcal{F}_t$ for all $t\geq 0$.   To each stopping time $T$, we can associate the optional $\sigma$-algebra
\begin{equation}
\mathcal{F}_T = \left\{\Phi 
\in \mathcal{F}: \Phi \cap \left\{\omega\in \Omega:T(\omega)\leq t\right\}\in \mathcal{F}_t, \: \forall t\right\}, 
\end{equation} 
which is a sub-$\sigma$-algebra of $\mathcal{F}$.   The entropy production $S(T)$ is defined  through the Radon-Nikodym derivative process 
\begin{equation}
\exp(-S(T)) = \frac{d\mathbb{P}^\dagger|_{\mathcal{F}_T}}{d\mathbb{P}|_{\mathcal{F}_T}}, \label{eq:STProb}
\end{equation}
where now $\left.\mathbb{P}\right|_{\mathcal{F}_T}$ and $\left.\mathbb{P}^\dagger\right|_{\mathcal{F}_T}$ are the restrictions of the measures $\mathbb{P}$ and $\mathbb{P}^\dagger$ to the $\sigma$-algebra $\mathcal{F}_T$.
The average entropy production $\langle S(T)\rangle$ takes thus the form 
 \begin{equation}
 \langle S(T)\rangle = \int_{\omega\in \Omega} {\rm d} \mathbb{P}  \ln  \frac{d\mathbb{P}|_{\mathcal{F}_T}}{d\mathbb{P}^\dagger|_{\mathcal{F}_T}}(\omega). \label{eq:STApp}
 \end{equation}
 
In the case of a Markov chain the definitions of $\exp(-S(t))$ and $\exp(-S(T))$ as defined here correspond with   those  in the main text.   Notably, in this case we get that 
\begin{equation}
\exp(-S(t))  = \frac{p^\dagger(X^t_0)}{p(X^t_0)},
\end{equation}
with $p$  as defined in Eqs.~(\ref{eq:pPathDiscrete}) and (\ref{eq:PathContinuous}) in discrete and continuous time, respectively, and with  $p^\dagger$  the corresponding quantity for the time-reversed process determined by the matrix $\mathbf{q}^\dagger$.   Analogously, it holds that 
\begin{equation}
\exp(-S(T))  = \frac{p^\dagger(X^T_0)}{p(X^T_0)}.
\end{equation}
Note that the formula (\ref{eq:STProb}) also applies to processes for which the Radon-Nikodym derivative cannot be straightforwardly expressed as the ratio of two path probability densities, as in Langevin processes with multiplicative noise~\cite{roldan2022martingales}.

\subsection{Coarse-graining of  $\langle S(T)\rangle$}
We coarse-grain  the quantity  $\langle S(T)\rangle $   by defining a random variable, say $Y\in \mathcal{Y}$ that is $\mathcal{F}_T$-measurable, where $\mathcal{Y}$  is the set of possible values of $Y$.      From the definition of a  random variable it follows that it partitions the set $
 \Omega$  as 
 \begin{equation}
 \Omega =  \cup_{y\in \mathcal{Y}} \Phi_y
 \end{equation}
 where
 \begin{equation}
 \Phi_y = \left\{\omega \in \Omega: Y(\omega) = y\right\}. 
 \end{equation}
 Furthermore, since $Y$ is  $\mathcal{F}_T$-measurable, it holds that 
$\Phi_y\in \mathcal{F}_T$.  Note that $\Phi_y\cap \Phi_{y'}=\phi$ for all $y\neq y'$.

To coarse-graine $\langle S(T)\rangle$ with respect to the random variable $Y$, we follow a derivation similar to  Lemma 3.2.1 of Ref.~\cite{tartakovsky2014sequential}.  Since the $\Phi_y$ form  a partition of $\Omega$, we can decompose the right-hand side of Eq.~(\ref{eq:STApp}) as
\begin{equation}
\int_{\omega\in\Omega} {\rm d} \mathbb{P}\ln    \frac{d\mathbb{P}|_{\mathcal{F}_T}}{d\mathbb{P}^\dagger|_{\mathcal{F}_T}}(\omega) = \sum_{y\in \mathcal{Y}} \int_{\omega\in\Phi_y} {\rm d}\mathbb{P}  \ln  \frac{d\mathbb{P}|_{\mathcal{F}_T}}{d\mathbb{P}^\dagger|_{\mathcal{F}_T}}(\omega)  . \label{eq:A10}
\end{equation}
As  $\Phi_y\in \mathcal{F}_T$ and  $ \ln  \frac{d\mathbb{P}|_{\mathcal{F}_T}}{d\mathbb{P}^\dagger|_{\mathcal{F}_T}} $ is a $\mathcal{F}_T$-measurable, we can express the right-hand side of the previous equation by 
\begin{equation}
\sum_{y\in \mathcal{Y}} \int_{\omega\in\Phi_y} {\rm d} \mathbb{P}  \ln  \frac{d\mathbb{P}|_{\mathcal{F}_T}}{d\mathbb{P}^\dagger|_{\mathcal{F}_T}}(\omega)= \sum_{y\in \mathcal{Y}} \int_{\omega\in\Phi_y}\left.  {\rm d}\mathbb{P}\right|_{\mathcal{F}_T}(\omega)  \ln  \frac{d\mathbb{P}|_{\mathcal{F}_T}}{d\mathbb{P}^\dagger|_{\mathcal{F}_T}}(\omega). \label{eq:integral}
\end{equation}
Notice that the left-hand side of (\ref{eq:integral}) contains an integral over the probability space $(\Omega,\mathcal{F},\mathbb{P})$ whereas the right-hand side is an integral over the  probability space $(\Omega,\mathcal{F}_T,\left.\mathbb{P}\right|_{\mathcal{F}_T})$~\cite{tao2011introduction}. As $\ln$ is a strictly concave function, we can apply  Jensen's inequality 
to the expectation value with respect to the probability measure  $\mathbb{P}|_{\mathcal{F}_T}(\cdot)/\mathbb{P}|_{\mathcal{F}_T}(\Phi_y)$~\cite{Jensen1906, Durrett2019-os}  to get 
\begin{eqnarray}
 \int_{\omega\in\Phi_y} {\rm d} \mathbb{P}\vert_{\mathcal{F}_T}(\omega) \ln \frac{{\rm d} \mathbb{P}\vert_{\mathcal{F}_T}}{{\rm d}\mathbb{P}^\dagger\vert_{\mathcal{F}_T}}(\omega)  &=&\mathbb{P}\vert_{\mathcal{F}_T}(\Phi_y)  \int_{\omega\in\Phi_y}  \frac{{\rm d} \mathbb{P}\vert_{\mathcal{F}_T}}{\mathbb{P}\vert_{\mathcal{F}_T}(\Phi_y)}\ln \frac{{\rm d} \mathbb{P}\vert_{\mathcal{F}_T}}{{\rm d} \mathbb{P}^\dagger\vert_{\mathcal{F}_T}}(\omega)  \nonumber\\ 
 &\geq&  \mathbb{P}\vert_{\mathcal{F}_T}(\Phi_y)\ln \frac{\mathbb{P}\vert_{\mathcal{F}_T}(\Phi_y)}{\mathbb{P}^\dagger\vert_{\mathcal{F}_T}(\Phi_y)} .   \label{eq:A12}
\end{eqnarray}

Combining Eq.~(\ref{eq:A10}) with (\ref{eq:integral}) and (\ref{eq:A12}) we obtain 
\begin{equation}
\langle S(T)\rangle \geq \sum_{y\in \mathcal{Y}} \mathbb{P}\vert_{\mathcal{F}_T}(\Phi_y)\ln \frac{\mathbb{P}\vert_{\mathcal{F}_T}(\Phi_y)}{\mathbb{P}^\dagger\vert_{\mathcal{F}_T}(\Phi_y)} , \label{eq:final}
\end{equation}
which  is the final result. 
The equality in Eq.~(\ref{eq:final}) is attained when $ \ln \frac{d\mathbb{P}|_{\mathcal{F}_T}}{d\mathbb{P}^\dagger|_{\mathcal{F}_T}}$ is constant almost everywhere on $\Phi_y$.   
 Note that  Eq.~(\ref{eq:SIntermediate}) in  Sec.~\ref{sec:coarseGrainST} is a special case of (\ref{eq:final}) when $Y=D$.  In case of $Y=(D,T)$, the sum in Eq.~(\ref{eq:final})  becomes an integral and we recover Eq.~(\ref{eq:refinedIneq}).

\section{Large deviation theory for  first-passage times}\label{app:B}
We develop the large deviation theory for the first-passage problem $T$ defined in Eq.~(\ref{eq:T}).     Specifically, we derive the Eqs.~(\ref{eq:pM}), (\ref{eq:EADef}) and (\ref{eq:11}) from the main text that relate the large deviation properties of $T$ to those of $J$.

The tools we use to derive the large deviation properties of $T$ are based on martingale theory, see Refs.~\cite{raghu2024effective,neri2025martingale}.   Specifically, we use that the process 
\begin{equation}
M(t) = \phi_a(X(t)) \exp\left(-aJ(t) - \lambda_J(a)t\right) \label{eq:mart}
\end{equation}
 is   a martingale, where $\phi_a$ is the right eigenvector associated with the Perron root of $\tilde{\mathbf{q}}(a)$.   Applying Doob's optional stopping theorem to $M(T)$~\cite{williams1991probability,liptser2001statistics,roldan2022martingales}, we can relate the splitting probability $p_-$ and the moment generating functions $m_-$ and $m_+$ in the limit of large thresholds to the scale cumulant generating function $\lambda_J$.   These arguments date back to the seminal works of  Ville~\cite{Ville} and  Wald~\cite{Wald2, Wald1}, who analysed the first-passage problem (\ref{eq:T})    in the case when  $J(t)$ is a sum of independent and identically distributed random variables.   In this case,  the martingale $M(t)$ is identical to 
 Wald's martingale~\cite{Wald2,asmussen2010ruin} for which $\phi_a=1$.  In the present context of  time-additive observables in Markov processes, martingales of the form (\ref{eq:mart}) have   been studied in Refs.~\cite{raghu2024effective,neri2025martingale}  and similar martingales  appear in the theory of Markov additive processes, see~Refs.~\cite{asmussen2000multi,asmussen2010ruin} and references therein.

 Note that the martingales (\ref{eq:mart}) apply to both continuous-time and discrete-time Markov chains.   The distinction between both cases is in the definition of the tilted matrix $\tilde{\mathbf{q}}(a)$.  In continuous time, the $\mathbf{q}$-matrix in Eq.~(\ref{eq:qtilde}) is the transition rate matrix of a continuous-time Markov chain, and in discrete time the  $\mathbf{q}$-matrix in Eq.~(\ref{eq:qtilde}) is the matrix of transition probabilities between states.  Consequently, in continuous time $\lambda_J(a)$ is the Perron root of $\tilde{\mathbf{q}}(a)$, while in discrete time $\lambda_J(a)$ is the logarithm of the Perron root of  $\tilde{\mathbf{q}}(a)$. 

 This Appendix is structured into two parts.   In the first~\ref{app:B1} we investigate the decay of the splitting probability $p_-$ with the threshold parameter $\ell_-$, and in ~\ref{app:B2} we focus on the moment generating functions of $T$ conditioned on hitting the positive (or negative) threshold first.

\subsection{Splitting probability $p_-$}\label{app:B1}
We  show that  for $\overline{j}>0$ the splitting probability $p_-$ is an exponentially decaying function of $\ell_-$, i.e., 
\begin{equation}
\lim_{\ell_-\rightarrow \infty} \frac{|\ln p_-|}{\ell_-} = a^\ast. \label{eq:split}
\end{equation}
Additionally, we show that  $a^\ast$ is the nonzero solution of 
\begin{equation}
\lambda_J(a^\ast) = 0, \label{eq:lambdaJDefa}
\end{equation}
with $\lambda_J$ the scaled cumulant generating function of $J$.    Note that in Eq.~(\ref{eq:split}) the threshold $\ell_+$ can take an arbitrary finite value, or can even  be an arbitrary function of~$\ell_-$.     

We derive the aforementioned results following calculations that are 
 similar to the ones in Ref.~\cite{raghu2024effective}.  However, the derivation here  applies to both discrete-time and continuous-time Markov chains, while  Ref.~\cite{raghu2024effective} considered continuous-time Markov chains.  
 
As $\mathbb{P}(T<\infty)=1$ and $M(t)\in (-\ell_-,\ell_+)$ for $t<T$, we can apply Doob's optional stopping theorem~\cite{williams1991probability,liptser2001statistics,roldan2022martingales}
\begin{equation}
\langle M(T)\rangle = \langle M(0) \rangle = \langle \phi_a(X(0))\rangle
\end{equation}
to the martingale $M_t$.   Setting  $a=a^\ast$, where $a^\ast$ is defined as the nonzero solution to Eq.~(\ref{eq:lambdaJDefa}), we recover the  identity
\begin{equation}
    p_- \langle  \phi_{a^\ast}(X_T)\rangle_- e^{a^\ast\ell_- (1+o_{\ell_-}(1))} + p_+  \langle \phi_{a^\ast}(X(T)) e^{-a^\ast  J(T) } \rangle_+= \langle \phi_{a^\ast}(X_0)\rangle. \label{eq:doob1}
\end{equation}
Here, $\langle \cdot \rangle_- = \langle \cdot | J(T)\leq -\ell_-\rangle $ is the average  conditioned on crossing the negative threshold first, and $\langle \cdot \rangle_+ = \langle \cdot | J(T)\geq \ell_+\rangle $ is conditioned on crossing the positive threshold first.   In the first term of (\ref{eq:doob1}) we have used that $J(T) = -\ell_- (1+o_{\ell_-}(1))$ where $o_{\ell_-}(1)$ represents an arbitrary function that converges to zero for large values of $\ell_-$.    Using 
\begin{equation}
\mathbb{P}(T<\infty) = p_- + p_+  =1,
\end{equation}
we can  substitute the $p_+$  in  Eq.~(\ref{eq:doob1})  by $1-p_-$.  Next, taking the limit $\ell_-\rightarrow \infty$ and using that the set $\mathcal{X}$ is finite so that  $|\phi_a(x)|$ can be bounded from above, we   recover the Eq.~(\ref{eq:split}) with $a^\ast$ the solution to  (\ref{eq:lambdaJDefa}).

\subsection{Generating functions $m_+$ and $m_-$}\label{app:B2}
We show  that the generating functions 
\begin{equation}
m_+(\mu) = \lim_{\ell_{+}\rightarrow \infty} \frac{1}{\ell_+}\ln \langle e^{\mu T}\rangle_+ \quad {\rm and} \quad m_-(\mu) = \lim_{\ell_{-}\rightarrow \infty} \frac{1}{\ell_-}\ln \langle e^{\mu T}\rangle_-  \label{eq:muPmuMv2}
\end{equation}
are for $\overline{j}>0$  the functional inverse of the scaled cumulant generating function $\lambda_J(a)$~\cite{budini2014fluctuating,gingrich2017fundamenta, neri2025martingale}, viz.,
\begin{equation}
\lambda_J(m_+(\mu)) = -\mu  \quad {\rm and} \quad \lambda_J(a^\ast-m_-(\mu)) = -\mu. \label{eq:muPmuM}
\end{equation}
Equation~(\ref{eq:muPmuM}) implies that the conditional distributions $p_T(t|+)$ and $p_T(t|-)$ satisfy a large deviation principle at both thresholds.  

Note that  Refs.~\cite{budini2014fluctuating,gingrich2017fundamenta}  considers a first-passage problem with one threshold, while here we consider  a first-passage problem with two thresholds~\cite{neri2025martingale}.    In the latter case, for $m_-$ the threshold $\ell_+$ can take an arbitrary finite value, or can even be an arbitrary function of $\ell_-$, and the converse is  true for $m_+$.   The implication is that  the rate function $\mathcal{I}_+$ is independent of the threshold $\ell_-$, and analogously, the rate function $\mathcal{I}_-$ is independent of $\ell_+$.   

As in \ref{app:B1}, we use  martingale arguments, this time following  Ref.~\cite{neri2025martingale}. However, the derivation here applies to both continuous and discrete-time Markov chains, while in previous work the continuous time was considered.

 Applying Doob's optional stopping theorem to $M(T)$, we find 
\begin{align}
    \langle M(T)\rangle &= p_+\langle\phi_a(X_T) \exp(-aJ(t) - \lambda_J(a)T)\rangle_+  \nonumber \\ 
    &+p_-\langle\phi_a(X_T) \exp(-aJ(t) - \lambda_J(a)T)\rangle_- = \langle \phi_a(X(0))\rangle .  \label{eq:Doob}
    \end{align}
We take the  parameter $a$ to be a function of $\mu$ so that 
\begin{equation}
\lambda_J(a) = -\mu. \label{eq:defAs}
\end{equation}
For a fluctuating current $J(t)$, this equation has two solutions, which we denote by  $a_+(\mu)>-a_-(\mu)$. 

For $a=a_+(\mu)$, we can express (\ref{eq:Doob}) as   
\begin{align}
p_+ e^{-a_+(\mu)\ell_+[1+o_{\ell_+}(1)]}  \langle e^{\mu T}\rangle_+ =  \langle \phi_{a_+}(X(0))\rangle  - p_- \langle\phi_{a_+}(X(T)) e^{-a_+(\mu)J(T) + \mu T}\rangle_-,
    \end{align}
    where we have used  that $J(T) = \ell_+ (1+o_{\ell_+}(1))$ and that $|\phi_a(x)|$ is bounded from above in the case of finite sets $|\mathcal{X}|$.    Taking the logarithm of both sides  and using the definition of $m_+(\mu)$ in Eq.~(\ref{eq:muPmuMv2}), we obtain   in the limit of large $\ell_+$ the identity  
    \begin{equation}
  m_+(\mu)  =  a_+(\mu) (1+o_{\ell_+}(1)). \label{eq:mPaP}
    \end{equation}
    Plugging (\ref{eq:mPaP}) into the Eq.~(\ref{eq:defAs}) for the definition of $a_+$, we recover the first equality in Eq.~(\ref{eq:muPmuM}).

Instead, for $a=-a_-(\mu)$, we can express (\ref{eq:Doob}) as  
    \begin{align}
p_- e^{-a_-(\mu)\ell_-[1+o_{\ell_-}(1)]}  \langle e^{\mu T}\rangle_- =  \langle \phi_{-a_-}(X(0))\rangle  - p_+ \langle\phi_{-a_-}(X(T)) e^{a_-(\mu)J(T) + \mu T}\rangle_+. 
    \end{align}
Taking the logarithm of both sides, we obtain in the limit of large threshold values $\ell_-$  the identity  
    \begin{align}
    -a_-(\mu) = (a^\ast - m_-(\mu))(1+o_{\ell_-}(1))  \label{eq:aMApp}
    \end{align}
    where we have used  Eq.~(\ref{eq:split}) for  $p_-$  and the Eq.~(\ref{eq:muPmuMv2}) for the definition of $m_-$.  Plugging (\ref{eq:aMApp}) into the Eq.~(\ref{eq:defAs}) for the definition of $-a_-$, we recover the second equality in Eq.~(\ref{eq:muPmuM}).

\section{Wald's equation for the fluctuating entropy production}\label{app:E}
We derive the following asymptotic version of Wald's equation 
\begin{equation}
\langle S(T)\rangle = \dot{s}\langle T\rangle (1+o_{\ell_{\rm min}}(1)).\label{eq:STAvAlb}
\end{equation}
Here, $S$ is the fluctuating entropy production, as defined in  Eq.~(\ref{eq:StDef}), and   $T$ is the first-passage time of a fluctuating current, as defined in Eq.~(\ref{eq:T}).   The Eq.~(\ref{eq:STAvAlb}) applies to 
 processes $X$ that are ergodic.  

\subsection{Wald's equation}
We use Wald's equation for sums of independent and identically distributed random variables~\cite{Wald2,Wald1,blackwell1946equation}.   Let $Y_1,Y_2,\ldots$ be an infinite sequence of independent and identically distributed random variables.     Consider the sum 
\begin{equation}
L_N = \sum^N_{j=1} Y_j \label{eq:sum}
\end{equation}
with      $N$  a stopping time of the process $Y_j$. 
Then,   Wald's equation~\cite{Wald2,Wald1,blackwell1946equation} 
\begin{equation}
\langle L_N \rangle = \langle N\rangle \langle Y_1\rangle \label{eq:Wald}
\end{equation}
holds if $\langle N \rangle<\infty$ and $\langle Y_1\rangle<\infty$.   Note that for Wald's equation it is not required that the $Y_j$ in the sequence $(Y_1,Y_2,\ldots,Y_N)$ are independent, but rather that the $Y_j$ in  $(Y_1,Y_2,\ldots,Y_n)$ for fixed $n\in \mathbb{N}$ are independent.   The condition $\langle N\rangle<\infty$ is essential.   For example, Wald's equation does not apply to a simple random walker on $\mathbb{Z}$ with one  absorbing boundary.   

To derive (\ref{eq:STAvAlb}), we need a generalised form of Wald's equation (see Theorem 12.9 in \cite{kallenberg}).  In this generalised form, we consider a stopping time $N$ of a  stochastic process $(X_1,X_2,\ldots)$, and a sequence $Y_1,Y_2,\ldots$  of independent and identically random variables for which $Y_j$ is a function of $X^j_1$.    Wald's equation (\ref{eq:Wald})  then also applies to the sum (\ref{eq:sum}) if $\langle N \rangle<\infty$ and $\langle Y_1\rangle<\infty$.

\subsection{Derivation of Eq.~(\ref{eq:STAvAlb})}

We derive Eq.~(\ref{eq:STAvAlb}) by using Wald's equation together with the fact that $X(t)$ is  ergodic  on a finite set $\mathcal{X}$.   

As $S(t)$ is a time-additive observable, we can decompose it as 
\begin{equation}
S(t) = S_r(t)  + \sum^{N(t)}_{j=1}\Delta S_j  \label{eq:S1}
\end{equation}
where $\Delta S_j = S(T_j)-S(T_{j-1})$, $S_r = S(t)-S(T_{N(t)})$, $T_j$ is the $j$-th time that the process $X(t)$ returns to its initial state $X(0)$ (i.e.,  $X(T_j)=X(0)$ but $\lim_{\epsilon\rightarrow 0^+}X(T_j-\epsilon)\neq X(0)$), and $N(t)$ is the number of times that $X(t)$ has returned to the state $X(0)$ in the time interval $[0,t]$.   Notice that $T_0=0$.  

As $X(t)$ is a Markov process defined on a finite set $\mathcal{X}$, it satisfies the strong Markov property, and therefore the  $\Delta S_j$ are independent and identically distributed random variables~\cite{bremaud2013markov}.  Hence, we can apply Wald's equation to the sum in (\ref{eq:S1}) yielding 
\begin{equation}
\langle S(T)\rangle = \langle S_r\rangle + \langle N(T)\rangle \langle \Delta S_1\rangle. \label{eq:STAv}
\end{equation}

Analogously, we can decompose the stopping time $T$ as  
\begin{equation}
T=  T_r + \sum^{N(T)}_{j=1}\Delta T_j , \label{eq:C6}
\end{equation}
with $\Delta T_j = T_j-T_{j-1}$ and $T_r = T-T_j$.  Due to the strong Markov property the $\Delta T_j$ are independent, and hence we can apply  Wald's equation to $T$ to obtain
\begin{equation}
\langle T\rangle = \langle T_r\rangle + \langle N(T)\rangle \langle \Delta T_1\rangle.   \label{eq:NT}
\end{equation}
Solving (\ref{eq:NT}) towards $\langle N(T)\rangle$ and substitution in (\ref{eq:STAv}) yields
\begin{equation}
\langle S(T)\rangle = \langle S_{r}(T)\rangle + (\langle T\rangle-\langle T_r\rangle) \frac{\langle \Delta S_1\rangle}{\langle \Delta T_1\rangle}. \label{eq:STAv2}
\end{equation}

Next, we use the fact that $X(t)$ is an ergodic process.  This property implies that the number of returns $N(t)$ to the initial state diverges as $t\rightarrow \infty$.     Consequently, in the limit of $\ell_{\rm min}\rightarrow \infty$, the quantities $\langle N(T)\rangle$ and $\langle T\rangle$ diverge, while  $ \langle S_r\rangle \in o_{\ell_{\rm min}}(1)$ and $ \langle T_r\rangle \in o_{\ell_{\rm min}}(1)$.  Thus, we obtain that   
\begin{equation}
\langle S(T)\rangle = \langle T\rangle \frac{\langle \Delta S_1\rangle}{\langle \Delta T_1\rangle} (1+o_{\ell_{\rm min}}(1)). \label{eq:STAv3}
\end{equation}

Lastly, we use ergodicity to identify in Eq.~(\ref{eq:STAv3})
\begin{equation}
\dot{s} = \frac{\langle \Delta S_1\rangle}{\langle \Delta T_1\rangle},  \label{eq:tobeconfirmed}
\end{equation}
completing the derivation of (\ref{eq:STAvAlb}).  

We end this appendix with deriving     (\ref{eq:tobeconfirmed}).    As before, we decompose the numerator and denominator in the definition of $\dot{s}$, 
\begin{equation}
\dot{s} = \lim_{t\rightarrow \infty} \frac{\langle S(t)\rangle}{t},\label{eq:defsdot}
\end{equation}
in terms of returns to the initial state.  The numerator is given by  Eq.~(\ref{eq:S1}) and the denominator can be written as  
\begin{equation}
t  = T_r(t) _+ \sum^{N(t)}_{j=1}\Delta T_j,  
\end{equation}
which is similar to (\ref{eq:C6}).  
Using ergodicity and the  law of large numbers, in the limit of $t\gg 1$ we obtain  
\begin{equation}
S(t) =\langle N(t)\rangle \langle \Delta S_1\rangle (1+o_t(1))
\end{equation}
and 
\begin{equation}
t = \langle N(t)\rangle \langle \Delta T_1\rangle (1+o_t(1)), 
\end{equation}
which combined with the definition (\ref{eq:defsdot}) yields  (\ref{eq:tobeconfirmed}). 

\section{Numerical calculations and parameters for the example plots}\label{App:D}

We detail the numerical procedures used to generate the Figs.~\ref{fig:fig5} and \ref{fig:fig6}, and  we provide the used system parameters.  

\subsection{Two-dimensional random walker model --- Fig.\ref{fig:fig5}}\label{app:D1}
The functions $\hat{s}_{\rm FPR}$ and $\hat{s}_{\rm iFPR}$ in the Fig.~\ref{fig:fig5} are obtained from numerically calculating the  right-hand sides of the Eqs.~\eqref{eq:sjas} and \eqref{eq:newboundIJform}, as these are identical to the right-hand sides in the definitions \eqref{eq:sfpr} and \eqref{eq:sifpr}, respectively.    Thus, it is necessary   to determine $\dot{s}$, $\overline{j}$, $\tilde{a}$, and $\lambda_J$.

The $\mathbf{q}$-matrix is the one of a continuous-time random walker  on  a $3\times3$ lattice with periodic boundary conditions ($|\mathcal{X}|=9$). 
To determine $\dot{s}$ and $\overline{j}$,  the stationary distribution $p_{\rm ss}$  was obtained from  diagonalising  the  rate matrix $\mathbf{q}$.   To determine $\tilde{a}$ and $\lambda_J$,   the  tilted matrix $\tilde{\mathbf{q}}$  was constructed from $\mathbf{q}$ through the definition (\ref{eq:qtilde}) by setting $c_{xy}$  to one of the values $1+\Delta$, $1-\Delta$, $-1-\Delta$, or $-1+\Delta$, in accordance with the definition Eq.~(\ref{eq:JDef}) of $J(t)$. The scaled cumulant generating function $\lambda_J(a)$ was obtained from   diagonalising the  tilted matrix.
 The non-zero root $a^\ast$ of $\lambda_J(a)$, and the root $\tilde{a}$ of $\lambda_J'(a) - \overline{j}$  were numerically calculated.

\subsection{Four state Markov jump process -- Fig.~\ref{fig:fig6}}\label{app:D2}
For the four state jump process of Sec.~\ref{sec:fourstate}, the rate matrix $\mathbf{q}$ is a $4\times4$ matrix.    The rate matrix used for the plots in panels (c) and (d) of Fig~\ref{fig:fig6} is 
\begin{equation}
    \mathbf{q} = \begin{bmatrix}
-701.18382264 & 630.11131304 & 0.00000000 & 71.07250961 \\
0.06591070 & -73.07279367 & 41.13760189 & 31.86928109 \\
0.00000000 & 1.37263879 & -15.79800672 & 14.42536793 \\
0.28046954 & 0.60384782 & 2.37931835 & -3.26363570
\end{bmatrix}.
\end{equation}
The rate matrix used for the plots in panels (e) and (f) of Fig~\ref{fig:fig6} is 
\begin{equation}
\mathbf{q}=\begin{bmatrix}
-49.88608088 & 34.54237606 & 0.00000000 & 15.34370482 \\
0.55651439 & -204.92622772 & 83.33329486 & 121.03641847 \\
0.00000000 & 0.18220575 & -188.9075424 & 188.72533667 \\
1.29269972 & 0.17444825 & 0.05394782 & -1.52109579
\end{bmatrix}.\label{eq:d5}
\end{equation} 
These rate matrices were generated by setting 
\begin{equation}
    \mathbf{q}_{xy} = \mathbf{q}^0_{xy} \exp(w_{xy}),
\end{equation}
and by drawing  the  $\mathbf{q}^0_{xy}$ and $w_{xy}$    for all $x\neq y$ independently  from a uniform distribution on $[1,10)$.  The diagonal elements of the matrix $\mathbf{q}$ are  set to $\mathbf{q}_{xx} = -\sum_{y\in \mathcal{X}}\mathbf{q}_{xy}$ to ensure that $\mathbf{q}$ is a stochastic matrix.

 The current $J(t)$ is determined by the coefficients $c_{xy}$, which for a given value of the angle $\beta$ are set to
\begin{eqnarray}
    c_{21} = -{c_{12}} &=& c_{1} = \cos\left(\tan^{-1}\left(\frac{\overline{j}_2}{\overline{j}_1}\right)-\beta\right),\\
    c_{23} = -{c_{32}} &=& c_{2} = \sin\left(\tan^{-1}\left(\frac{\overline{j}_2}{\overline{j}_1}\right)-\beta\right),
\end{eqnarray}
and $c_{xy} =0$ for all other $x,y\in\mathcal{X}$. With these values of $c_{xy}$, the tilted matrix $\tilde{\mathbf{q}}$ is given by Eq.~\eqref{eq:qtilde}.  

The quantities $\hat{s}_{\rm FPR}$ and $\hat{s}_{\rm iFPR}$ are obtained   from calculating the right hand sides of the Eqs.~\eqref{eq:sjas} and \eqref{eq:newboundIJform} with the same procedure as explained in \ref{app:D1}.

\section{Currents belonging to $\mathcal{J}_{\rm v}$ but not $\mathcal{J}_{\rm sym}$}\label{app:F}
We present an example of a current that satisfies the  weak symmetry relation \eqref{eq:weaksymm}, but not the Galavotti-Cohen symmetry \eqref{eq:GCsymLDR}.   Thus, this current belongs to the set $\mathcal{J}_{\rm v}\setminus \mathcal{J}_{\rm sym}$ in the Venn diagram of Fig.~\ref{fig:fig3}.   The example is taken from  the four-state Markov jump process discussed in Sec.~\ref{sec:fourstate}, with the jump rates corresponding with those of  the plots (e) and (f) of Fig.~\ref{fig:fig6} (as specified in Eq.~\eqref{eq:d5}). 

\begin{figure}[!ht]
    \centering
    \includegraphics[width=0.8\linewidth]{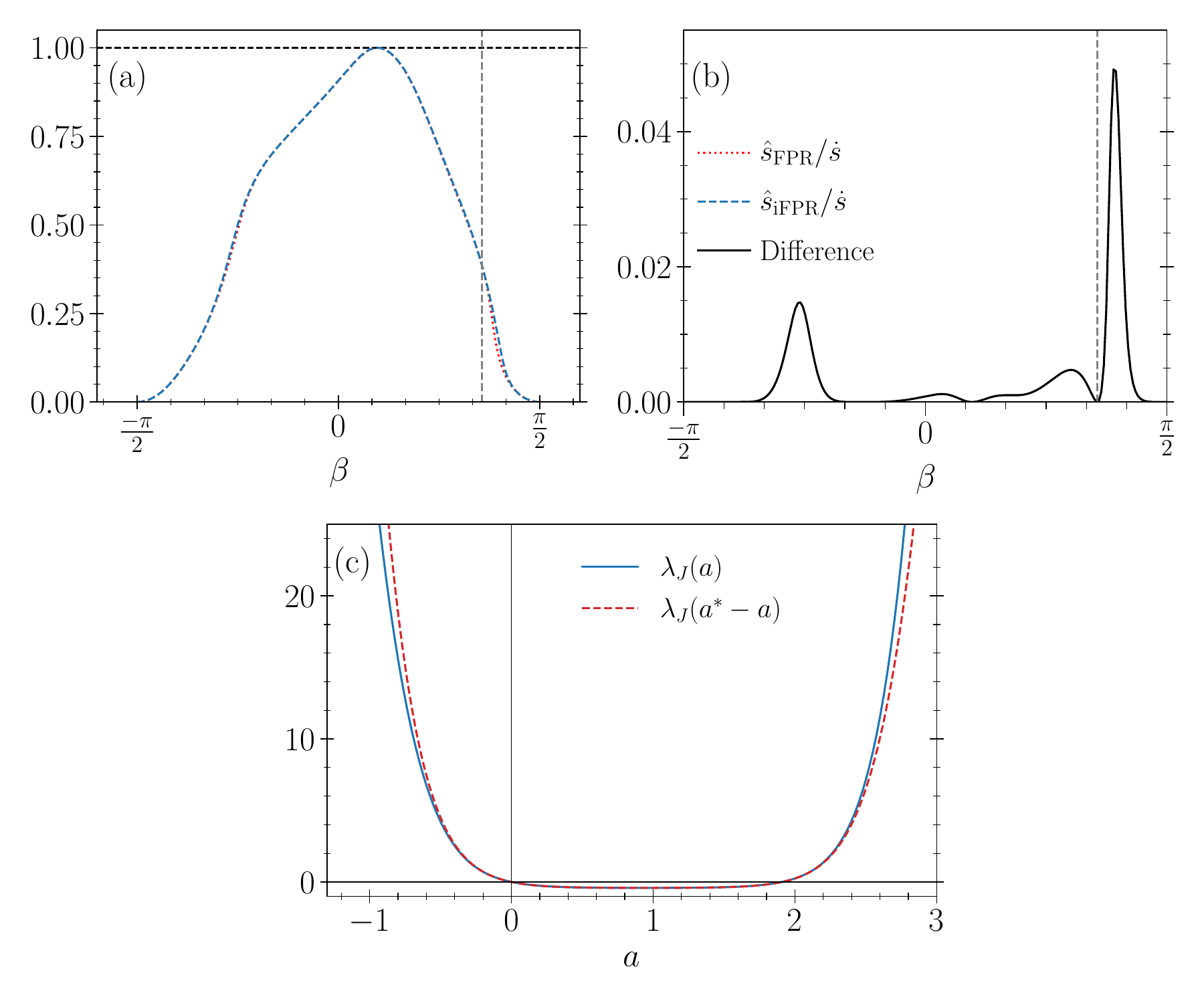}
    \caption{(a) Plot of $\hat{s}_{\rm FPR}/\dot{s}$ (red dotted line) and $\hat{s}_{\rm iFPR}/\dot{s}$ (blue dashed line) for the four state model illustrated in Fig.~\ref{fig:fig6}(a). The vertical grey dotted line indicates $\beta = 1.11844$, corresponding to currents $J(t)$ that satisfy the symmetry \eqref{eq:weaksymm}, but not the Galavotti-Cohen symmetry \eqref{eq:GCsymLDR}. (b) Plot of $(\hat{s}_{\rm FPR}-\hat{s}_{\rm iFPR})/\dot{s}$ (solid black line). The vertical grey line indicates $\beta = 1.11844$ where the difference vanishes and $\hat{s}_{\rm FPR}=\hat{s}_{\rm iFPR}$. This implies that the weak symmetry \eqref{eq:weaksymm} is satisfied (see Eq.\eqref{eq:weaksymproof}). (c) Plot of $\lambda_J(a)$ (solid blue line) and $\lambda_J(a^\ast - a)$ (dashed red line) for $\beta = 1.11844$. The Galavotti-Cohen symmetry \eqref{eq:GCsymlambda} is clearly violated.}
    \label{fig:fig7}
\end{figure}

The fluctuation and symmetry properties of a current $J$ in this model are determined by the parameter $\beta$ as defined in Sec.~\ref{sec:fourstate}, as the parameter $\beta$ determines the scaled cumulant generating function $\lambda_J(a)$. We find a value of  $\beta$ for which  $\hat{s}_{\rm FPR}$ equals $\hat{s}_{\rm iFPR}$  and for which the current is not optimal.   In the present example we find  with high numerical accuracy a value of  $\beta$ that is approximately equal to   $1.11844$ (see Panels (a) and (b) in Fig.~\ref{fig:fig7})  by numerically solving the equation $\hat{s}_{\rm FPR}-\hat{s}_{\rm iFPR}=0$ for $\beta \in (\pi/3 ,9\pi/24)$.     From Eqs.~\eqref{eq:sfpr}~and~\eqref{eq:sifpr} it follows that for  currents corresponding with this value of $\beta$ the weak symmetry \eqref{eq:weaksymm} is satisfied, as
\begin{equation}
    \mathcal{I}_- (1/\overline{j}) = 0 \implies \lim_{\ell_- \to \infty} \frac{\langle T \rangle_-}{\ell_-} = \frac{1}{\overline{j}} = \lim_{\ell_+ \to \infty} \frac{\langle T \rangle_+}{\ell_+}. \label{eq:weaksymproof}
\end{equation}

Furthermore, we show that currents corresponding to $\beta \approx 1.11844$. do not satisfy the  Galavotti-Cohen symmetry \eqref{eq:GCsymLDR}.   The Galavotti-Cohen symmetry can be expressed in terms of the scaled cumulant generating function $\lambda_J(a)$ of the current as 
\begin{equation}
    \lambda_J(a) = \lambda_J(a^\ast -a),\label{eq:GCsymlambda}
\end{equation}
where we recall that $a^\ast$ is the effective affinity.  Panel (c) of Fig.~\ref{fig:fig7} shows that the equation (\ref{eq:GCsymlambda}) is not satisfied for  $\beta \approx 1.11844$.  

\end{document}